\documentclass[sigconf,final]{acmart}

\usepackage{etoolbox,siunitx}
\robustify\bfseries

\AtBeginDocument{%
  \providecommand\BibTeX{{%
    \normalfont B\kern-0.5em{\scshape i\kern-0.25em b}\kern-0.8em\TeX}}}

\setcopyright{acmcopyright}
\copyrightyear{2020}
\acmYear{2020}
\acmDOI{10.1145/1122445.1122456} 

\acmConference[ICMI '20]{ICMI '20: ACM International Conference on Multimodal Interaction}{October 25--29, 2020}{Utrecht, the Netherlands}
\acmBooktitle{ICMI '20: ACM International Conference on Multimodal Interaction,
  October 25--29, 2020, Utrecht, the Netherlands}
\acmPrice{15.00} 
\acmISBN{978-1-4503-XXXX-X/18/06} 
 
\acmSubmissionID{1303}

\begin{document}

\title{Toward Multimodal Modeling of Emotional Expressiveness}

\author{Victoria Lin}
\authornote{Both authors contributed equally to this research.}
\orcid{0000-0003-2873-2177}
\affiliation{\institution{Carnegie Mellon University}}
\email{vlin2@andrew.cmu.edu}

\author{Jeffrey M. Girard}
\authornotemark[1]
\orcid{0000-0002-7359-3746}
\affiliation{\institution{Carnegie Mellon University}}
\email{jmgirard@cmu.edu}

\author{Michael A. Sayette}
\orcid{0000-0001-7617-5198}
\affiliation{\institution{University of Pittsburgh}}
\email{sayette@pitt.edu}

\author{Louis-Philippe Morency}
\orcid{0000-0001-6376-7696}
\affiliation{\institution{Carnegie Mellon University}}
\email{morency@cs.cmu.edu}

\renewcommand{\shortauthors}{Lin, Girard, Sayette, and Morency}

\begin{abstract}
 Emotional expressiveness captures the extent to which a person tends to outwardly display their emotions through behavior. Due to the close relationship between emotional expressiveness and behavioral health, as well as the crucial role that it plays in social interaction, the ability to automatically predict emotional expressiveness stands to spur advances in science, medicine, and industry. In this paper, we explore three related research questions. First, how well can emotional expressiveness be predicted from visual, linguistic, and multimodal behavioral signals? Second, how important is each behavioral modality to the prediction of emotional expressiveness? Third, which behavioral signals are reliably related to emotional expressiveness? To answer these questions, we add highly reliable transcripts and human ratings of perceived emotional expressiveness to an existing video database and use this data to train, validate, and test predictive models. Our best model shows promising predictive performance on this dataset ($RMSE=0.65$, $R^2=0.45$, $r=0.74$). Multimodal models tend to perform best overall, and models trained on the linguistic modality tend to outperform models trained on the visual modality. Finally, examination of our interpretable models' coefficients reveals a number of visual and linguistic behavioral signals---such as facial action unit intensity, overall word count, and use of words related to social processes---that reliably predict emotional expressiveness.
\end{abstract}



\maketitle

\section{Introduction}
\label{sec:intro}

\textit{Emotional expressiveness} is a psychological attribute that captures the degree to which a person tends to display their emotions through behavior \cite{kring1994,gross1995}. A highly expressive person tends to exhibit what they feel through their facial movements, vocalizations, and other gestures, whereas a less expressive person tends to hide or hold their feelings in. 
Emotional expressiveness is related to but distinct from other affect-related traits. It is much narrower in its scope (i.e., more specific in terms of the patterns of behavior it explains) than broad personality traits such as extraversion and neuroticism \cite{mccrae1999}, and it differs from emotionality \cite{tellegen1988} in that it describes the frequency and intensity of emotional \textit{displays} rather than the frequency and intensity of emotional \textit{experiences}. 

Greater scientific understanding of emotional expressiveness is critical for many areas of basic and applied research. For example, correctly inferring a person's mental state from their behavior (e.g., how much they are enjoying a comedy) likely requires careful consideration of how expressive that person is generally: a small chuckle from a highly expressive person may mean that they are unimpressed, whereas that same chuckle from a normally tight-lipped person may mean the opposite. Similarly, when synthesizing expressive behavior for a robot or virtual agent, it is important to consider how much emotion it should display over the course of an interaction. For example, there may be some settings where a more expressive robot or agent would increase user engagement and enjoyment \cite{hamacher2016believing} and others where such displays would be inappropriate or distracting. Users may also prefer a robot or agent that matches his or her level of expressiveness, which would require the ability to detect user expressiveness and then modulate the synthesis of behavior accordingly. Finally, emotional expressiveness plays an important role in healthcare, where persistent abnormalities in the display of emotion (e.g., reduced or inappropriate emotional expressiveness) serve as observable signs of various medical conditions (e.g., depression, mania, schizophrenia, and brain injury) \cite{americanpsychiatricassociation2013}. The ability to automatically detect such abnormalities could enhance the objectivity and efficiency of clinical assessment.

Furthermore, in many of these applications, interpretability is paramount. Understanding which behavioral signals are reliably related to emotional expressiveness would enable virtual agents and robots to more effectively simulate expressive behavior and would aid in the training of clinicians. Understanding which behavioral signals a model is using to predict emotional expressiveness would help to validate these models and would allow clinicians to discount their predictions when appropriate (e.g., when an important behavioral signal observed in a given patient is known to be caused by something other than disease, such as a drug side-effect).


In an ideal world, emotional expressiveness would be objectively measured as the \textit{true} frequency and intensity of emotional displays. However, the affective sciences currently lack a single, agreed-upon way of delineating and representing emotional displays \cite{barrett2019,cowen2019}, so this objective ground truth is currently out of reach. Instead, we focus on third-party observers' \textit{perceptual ratings} of emotional expressiveness. This approach has the potential to introduce subjectivity and bias, but we can quantify this subjectivity through inter-rater reliability analyses and mitigate it through the aggregation of ratings \cite{rosenthal1982,lin2019}. Operationalizing emotional expressiveness through observer ratings also has the benefit of paralleling the standard approach used in medicine, where clinicians observe patients' behavior and rate their emotional expressiveness \cite{americanpsychiatricassociation2013}.


Motivated by the potential applications described previously, the current study explores the following research questions:
\begin{enumerate}
    \item How well can we predict emotional expressiveness from multimodal representations of participants' behavior?
    \item How much do the different behavioral modalities contribute to the prediction of emotional expressiveness?
    \item Which behavioral signals are reliably related to emotional expressiveness? What is the nature of these relationships?
\end{enumerate}
To begin answering these questions, the current study makes three primary contributions to the literature. First, we add highly reliable transcripts and perceptual ratings of emotional expressiveness to an existing video database. Second, we train algorithms to predict emotional expressiveness from visual, linguistic, and multimodal representations of participants' behavior and compare their performance. Third, we explore what our best-performing algorithm learned about the interpretable relationships between behavioral signals and perceptual ratings of emotional expressiveness.

\section{Related Work}
\label{sec:literature}
Although computational work on emotional expressiveness has been rare, interdisciplinary researchers have extensively studied the adjacent topics of personality perception and emotion recognition. Both topics resemble the task of computationally characterizing emotional expressiveness, as they involve inferring psychological traits and states, respectively, from observed behavior. Therefore, we look to research in these areas to inform our own approach to predicting and characterizing emotional expressiveness. In this section, we describe the current state of these areas of study.

\subsection{Personality Perception}

In personality computing, the prediction of personality traits as rated by external observers is referred to as \emph{automatic personality perception} \cite{vinciarelli2014}. 
Studies in automatic personality computing typically classify whether an individual is above or below the median on a trait (although a more rigorous approach is to predict their dimensional scores \cite{decoster2009,wright2014b}). Previous models used acoustic \cite{polzehl2010automatically, schuller2012interspeech}, linguistic \cite{valente2012annotation}, and visual \cite{biel2012youtube} data, often in multimodal combinations. Traditional machine learning algorithms have been most common, especially support vector machines (SVMs) \cite{chastagnol2012personality, montacie2012pitch, staiano2011automatic}, although some interpretable approaches have been used  \cite{mohammadi2012automatic}.

Recently, deep learning approaches have played a larger role in personality computing. For example, in the ChaLearn LAP challenge \cite{ponce2016chalearn}, four models achieved competitive performance out of eight participating models. All four of these top-performing models used convolutional neural networks (CNNs) as feature extractors; to predict over these features, three models used deep learning regressors, and one used support vector regression (SVR).

\subsection{Emotion Recognition}

Generally, efforts in emotion recognition involve either the prediction of discrete emotion categories or continuous affective dimensions from short segments of behavior \cite{gunes2013a}. Notably, emotion recognition differs from both automatic personality perception and the emotional expressiveness prediction task in this paper in that it involves computing a momentary state rather than a longer-lasting trait. Nonetheless, emotion recognition shares many methodological similarities with automatic personality perception. 

Emotion recognition models have been trained on acoustic, linguistic, visual, and multimodal data \cite{zeng2009}. The most common approaches are machine learning classifiers including SVMs \cite{kapoor2005multimodal, vidrascu2005detection, schuller2007audiovisual}, deep belief networks \cite{cohen2003facial, tong2007facial}, and ensemble methods like AdaBoost \cite{littlewort2007faces, zeng2007audio, petridis2008audiovisual}. Recently, partially in response to emotion recognition challenges from video like EmotiW \cite{dhall2013emotion, dhall2015video, dhall2018emotiw} and AVEC \cite{valstar2014}, multimodal deep learning models---often comprised of CNN feature extractors over either a deep learning classifier \cite{tzirakis2017end} or a simpler method like SVM \cite{sharif2014cnn} or logistic regression \cite{donahue2014decaf}---have assumed a prominent role in emotion recognition tasks, particularly in the classification of basic emotion categories. Extensions of this type of architecture to multitask learning and cascade networks, as well as experiments with generative adversarial networks and variational autoencoders, produce the current state-of-the-art results \cite{li2020deep}.

\subsection{Emotional Expressiveness}
\label{sec:eelit}
Prior work in psychology focused on refining the definition and measurement of emotional expressiveness as a trait. Several questionnaires were developed to measure emotional expressiveness via self-report and were used to examine its relationship with other affective and personality attributes \cite{friedman1980,king1990,kring1994,gross1995}. These studies found that a general disposition to outwardly display emotions could be readily found and measured via self-report. This disposition was found to be most closely aligned with the personality trait of extraversion (i.e., sociability, assertiveness, and activity), although it was also related to neuroticism (i.e., anxiety, hostility, and impulsiveness) and agreeableness (i.e., trust, compliance, and modesty) in different contexts. Specifically, extraversion appears to be related to increased expression of both positive and negative emotion, whereas neuroticism appears to favor the expression of negative emotion, and agreeableness appears to favor the expression of positive emotion \cite{gross1995}. It is important to note, however, that these relationships were all modest in magnitude (e.g., correlations were around 0.2 to 0.4 with extraversion, --0.2 to 0.3 with neuroticism, and 0.0 to 0.2 with agreeableness), which supports the idea that emotional expressiveness is related to but still distinct from these broader personality traits.

Human observers have also been used to measure how emotionally expressive a person appears to be within a given interaction \cite{kring1994,sullivan2018}. However, these studies did not explore which specific behavioral signals the observers used to make their ratings. It is thus unclear which aspects of communicative behavior are most expressive. Recently, we sought to address this gap in the literature by automatically predicting facial expressiveness and quantifying its behavioral correlates on a moment-to-moment basis using the Elastic Net algorithm \cite{lin2019}. This work informs our choice of methods in the current paper but differs in a number of important ways. First, whereas the previous study examined expressiveness using only visual representations of behavior, the current study also examines linguistic and multimodal representations---both of which have been shown to be crucial in studies on personality computing and emotion recognition---in addition to visual representations. Second, whereas the previous study examined expressiveness on a momentary time-scale, the current study examines holistic patterns of expressiveness over a longer time period to better capture emotional expressiveness as an enduring pattern of behavior. Finally, whereas the previous study examined induced responses to emotional stimuli, the current study examines expressive behavior in a more naturalistic and unstructured social interaction setting.

\begin{figure}[t]
    \centering
    \includegraphics[width=\linewidth]{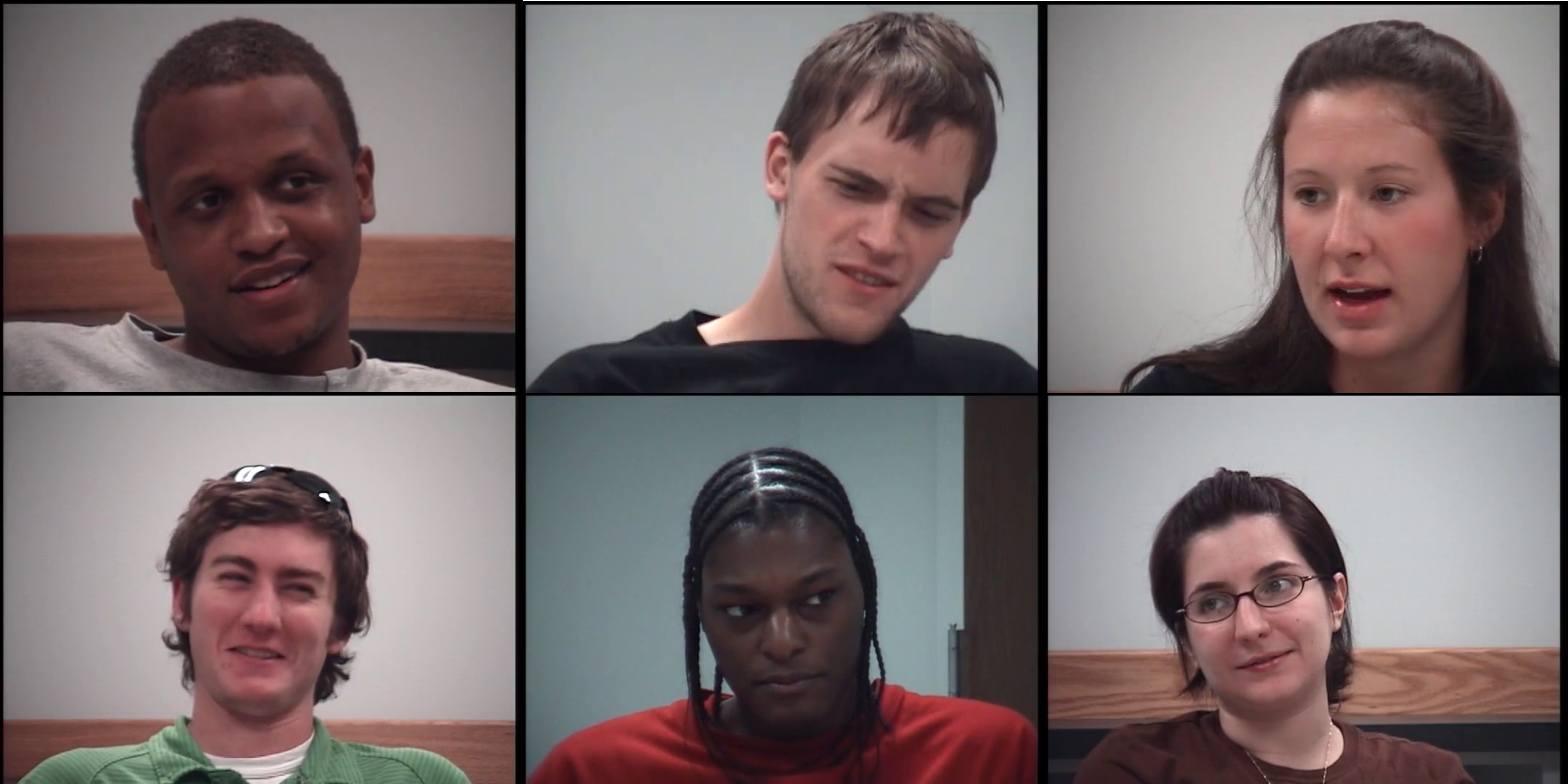}
    \caption{Example frames from the GFT database}
    \Description{}
    \label{fig:ex_frames}
\end{figure}

\section{Data}

The videos examined in this work came from the Sayette Group Formation Task (GFT) spontaneous facial expression database \cite{girard2017a}. This database includes videos of 96 participants (42\% Female, 85\% White) observed during a psychology experiment \cite{sayette2012}. In this experiment, a group of three participants would meet for the first time and get to know one another over the course of \SI{36}{\minute}; this time was unstructured, and groups could discuss whatever topics they wished. The GFT database includes \SI{1}{\minute} videos of each group, starting an average of \SI{5.6}{\minute} into the interaction. Separate wall-mounted cameras faced each participant and recorded a $720\times480$ pixel video of their head and shoulders at a rate of \SI{29.97}{fp\second}. Example frames from these videos can be found in \autoref{fig:ex_frames}.

Audio during the experiment was recorded from a single centrally located microphone.\footnote{The audio recordings and personality scores were not released as part of the official GFT database distribution \cite{girard2017a}. We were granted special access for this study.} In order to measure language-related behaviors, we manually transcribed text from the audio recordings. The speech of all 96 participants was transcribed by the first author and, for the purposes of assessing transcription reliability, the speech from a subset of 30 participants was also transcribed by the second author. Word-level agreement was high ($95.1$\%) on this subset of participants, which provides evidence that the transcriptions are trustworthy overall. 

Self-reported personality traits were measured using the NEO-FFI \cite{costa1992}, a 60-item inventory yielding scale scores on the ``Big Five'' personality traits: neuroticism, extraversion, openness to experience, agreeableness, and conscientiousness traits.\footnotemark[1] This inventory was completed by participants prior to the interaction.

\section{Measuring Expressiveness}
\label{sec:expressiveness}
In order to measure how emotionally expressive each participant in the GFT dataset was, we paid human annotators to watch and rate each video. To mitigate the influence of annotator subjectivity, each video was rated by multiple annotators, and ratings were averaged per video \cite{rosenthal1982}. Moreover, to increase the reliability and validity of these ratings, annotators completed multiple items (i.e., questions) measuring different aspects of emotional expressiveness, which we combined into a single score per video using latent variable modeling \cite{kline2015}. Although they take longer to complete, multiple-item scales are generally preferred to single-item scales because the latter tend to be quite noisy (i.e., lack reliability), struggle to capture the complexity and nuance of many attributes (i.e., lack validity), and can lead to incorrect analytical conclusions \cite{crede2012}. In this section, we describe the process of collecting these perceptual ratings and conducting the latent variable analysis. 

\subsection{Perceptual Ratings}
Human annotators were recruited through the \textit{Prolific.co} online research platform. In order to better assess annotator quality and consistency, we wanted each annotator to rate multiple videos. However, requiring each annotator to rate all the videos might lead to fatigue or dropout. To balance these two concerns, we separated the 96 participant videos into 6 sets of 16 videos and assigned a different group of annotators to rate each set. To minimize the influence of cultural and sensory differences, annotators were required to have USA nationality, English language fluency, and normal (or corrected to normal) vision and hearing. To ensure high-quality ratings, annotators were required to have a prior approval rate of 95\% or higher, and annotators were replaced if they failed one or more attention check questions during our experiment. For each set of videos, we recruited eight crowdworkers to serve as raters. After three annotators were replaced for failing attention checks, the final sample of 48 annotators was 63\% Female, 60\% White, and ranged in age from 19 to 60 years old ($M=30.2$, $SD=10.3$).

Using the \textit{formr} online survey platform \cite{arslan2020}, annotators viewed instructions and then rated each video in their assigned set. Annotators were instructed to watch each video (from start to finish at normal speed and with the volume on) and then carefully answer a series of 12 questions about how they perceived the person in the video. In this study, we focused on a subset of 4 of these questions that captured different aspects of emotional expressiveness. 

Although previous work described in \autoref{sec:eelit} developed methods for measuring emotional expressiveness, many of these self-report items were difficult to adapt to observational research (e.g., ``The way I feel is different from how others think I feel'') and previous observational studies used single-item measures (e.g., rating videos on ``expressiveness''), which we previously noted can attenuate reliability and validity. As such, we decided to design new questions to be used in a multiple-item scale. We chose four items to balance the competing needs for comprehensiveness (which requires more items) and efficiency (which requires fewer).

\begin{enumerate}
    \item How \textbf{expressive} was the person in this video (use your own understanding of what it means to be expressive)?
    \item How much did the person in this video show their \textbf{emotions} (through their words and nonverbal behavior)?
    \item How \textbf{animated} (lively, energetic, or active) was the person in this video?
    \item How much did the person in this video \textbf{react} to the other people (through their words and nonverbal behavior)?
\end{enumerate}

The first question represented the single-item measures used in previous observational studies, and the second question captured the theoretical definition of emotional expressiveness. The third and fourth questions were more exploratory and were our attempts to represent other observable aspects of emotional expressiveness. Specifically, the third question attempted to capture the idea that emotions are often strongly expressed through movement and activity, and the fourth question attempted to capture the idea that emotions (and therefore emotional expressions) are reactions to environmental (and often social) stimuli. All questions were carefully worded to generalize across all emotions and behavioral modalities. Each question was answered using a rating scale with five categorical options ranging from 0 (\textit{not at all}) to 4 (\textit{extremely}).

\begin{figure}[t]
    \centering
    \includegraphics[width=\linewidth]{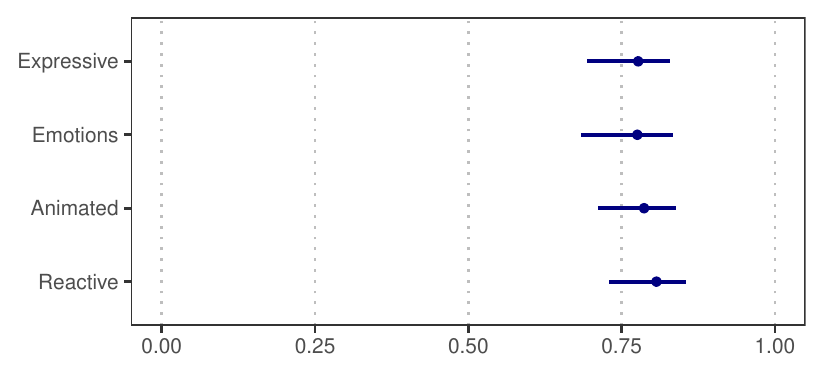}
    \caption{Inter-annotator reliability as measured by intraclass correlation coefficients (with 95\% confidence intervals)}
    \Description{}
    \label{fig:irr}
\end{figure}

The inter-annotator reliability of the perceptual ratings was calculated per question using the \textit{agreement} software package \cite{girard2020}. Intraclass correlation coefficients (ICCs) under model 2A \cite{gwet2014} were calculated to estimate the reliability of the average of all eight annotators' scores. ICC values above 0.50, 0.75, and 0.90 are generally considered to be evidence of moderate, good, and excellent reliability, respectively \cite{koo2016}. Note that these ICCs are more conservative than Pearson correlation coefficients \cite{mcgraw1996}. As shown in \autoref{fig:irr}, reliability estimates were all in the good range, and confidence intervals (CIs) did not extend below the moderate range.

\subsection{Latent Variable Modeling}
\label{sec:latent}
\begin{figure}[t]
    \centering
    \includegraphics[width=.75\linewidth]{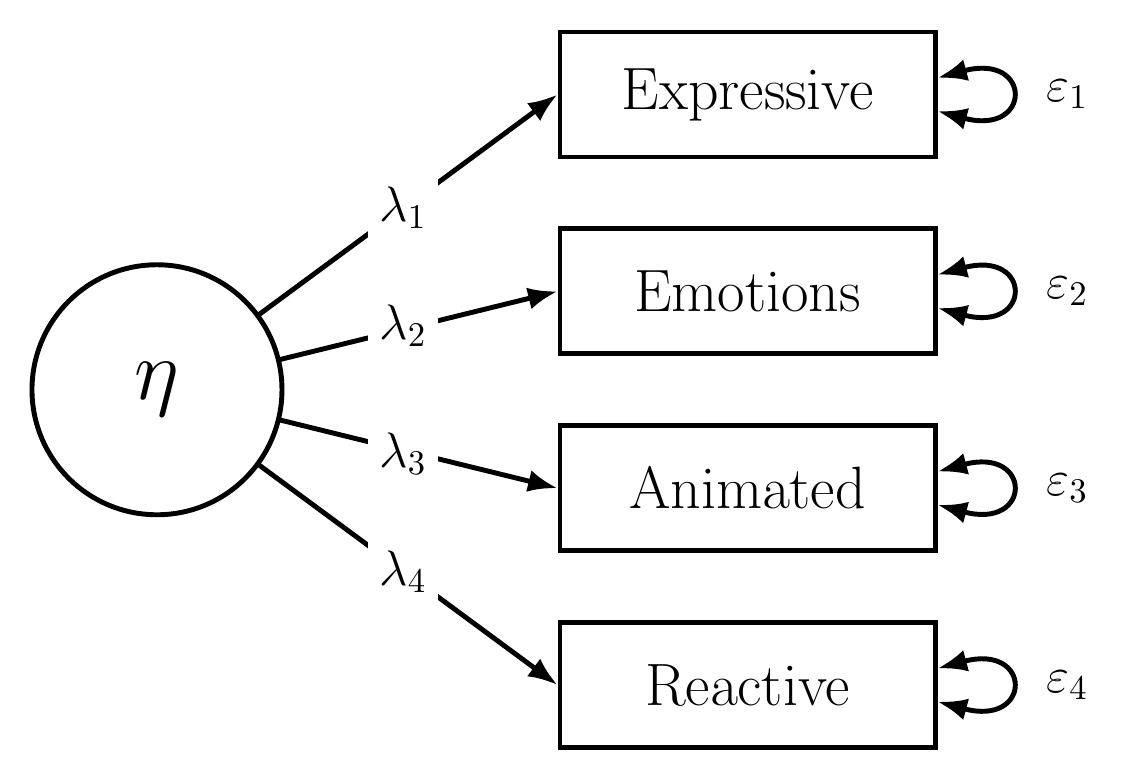}
    \caption{Path diagram of the CFA model (ovals are latent variables, rectangles are manifest/observed variables, $\protect\lambda$ are factor loadings, and $\protect\varepsilon$ are residual variances)}
    \label{fig:cfa}
\end{figure}

After being averaged across raters, scores on the four expressiveness-related questions were highly inter-correlated (all $r>0.83$), which supports the feasibility of combining these scores into a single lower-dimensional representation. The simplest way to combine the scores would have been to sum or average them, but to do so would assume that all questions are equally important and equally well-measured; these assumptions are unlikely to be met in practice and would negatively impact the validity and reliability of the aggregate if violated \cite{mcneish2020}. Thus, to avoid these assumptions, we used confirmatory factor analysis (CFA) \cite{kline2015} to estimate a latent variable $\eta$ that explains the variance shared among the questions (\autoref{fig:cfa}). In this model, factor loadings represent how much of each question was composed of shared variance (i.e., how strong the relationship was between the question and the latent variable), and residual variances represent how much each question was composed of non-shared variance (i.e., unique variance and measurement error). Because these model parameters are freely estimated (i.e., informed by the data), the latent variable $\eta$ will be a better representation of the shared variance than a sum or average would be \cite{mcneish2020}. 

The CFA model was estimated in a Bayesian framework using Markov Chain Monte Carlo in the \textit{blavaan} software package \cite{merkle2018}. All latent and manifest variables were standardized to zero mean and unit variance. Normal priors ($\mu=0$, $\sigma=1$) were used for the intercept and factor loading parameters, and gamma priors ($\alpha=1$, $\beta=1$) were used for the precision (i.e., residual standard deviation) parameters. Four Markov chains were estimated, each with 1000 warmup and 1000 inference iterations.

We evaluated the fit of the CFA model using Bayesian versions of two common fit indexes that are known to not be biased when calculated in small samples \cite{garnier-villarreal2020}: the gamma-hat fit index $(\hat{\Gamma})$, which is based on the noncentral chi-square distribution, and the comparative fit index (CFI), which compares the proposed model to both a baseline model (in which all covariances are set to zero) and a saturated model (in which all covariances are freely estimated). Both of these indexes range from 0 to 1 and are commonly interpreted as being indicative of good model fit when above 0.950 \cite{hu1999}.

The fit of the CFA model was good according to both fit indexes ($\hat{\Gamma}=0.981$, $\text{CFI}=0.993$). \autoref{tab:bcfa} provides a summary of the model's parameter estimates. All four questions had strongly positive loadings on the latent variable, which shows that they shared a great deal of variance (i.e., tended to vary together) and supports our argument that they measured a single underlying attribute \cite{flake2017}. 

\begin{table}[t]
    \caption{Results from the Bayesian CFA model}
    \label{tab:bcfa}
    \begin{tabular}{
        l
        l
        S[table-format = 1.2]
        >{{[}} 
        S[table-format = 2.2,table-space-text-pre={[}] 
        @{,\,} 
        S[table-format = 1.2,table-space-text-post={]}] 
        <{{]}} 
    }
    \toprule
    \multicolumn{2}{l}{Parameter} & {Estimate} & \multicolumn{2}{c}{95\% CI} \\
    \midrule
    $\lambda_1$ & Expressive Factor Loading & 0.97 & 0.84 & 1.14 \\
    $\lambda_2$ & Emotions Factor Loading   & 0.95 & 0.81 & 1.11 \\
    $\lambda_3$ & Animated Factor Loading   & 0.96 & 0.83 & 1.13 \\
    $\lambda_4$ & Reactive Factor Loading   & 0.87 & 0.72 & 1.05 \\
    \midrule
    $\varepsilon_1$ & Expressive Residual Variance   & 0.07 & 0.04 & 0.10 \\
    $\varepsilon_2$ & Emotions Residual Variance     & 0.11 & 0.07 & 0.16 \\
    $\varepsilon_3$ & Animated Residual Variance     & 0.08 & 0.05 & 0.12 \\
    $\varepsilon_4$ & Reactive Residual Variance     & 0.24 & 0.18 & 0.33 \\
    \bottomrule
    \end{tabular}
\end{table}

To further validate our latent variable as a measure of emotional expressiveness (as opposed to some other quantity), we extracted factor scores (i.e., estimates of each participant's standing on the latent variable $\eta$) and assessed their pattern of correlation with participants' self-reported personality traits. If our latent variable is a good estimate of emotional expressiveness, this pattern of correlation should match expectations based on theory and previous work \cite{flake2017}. Based on the literature reviewed in \autoref{sec:eelit}, we expected the factor scores to be moderately positively correlated with extraversion, weakly positively correlated with agreeableness, and weakly positively or negatively correlated with neuroticism depending on the negativity versus positivity of the context.

The factor scores were positively correlated with extraversion (0.26) and agreeableness (0.28) and slightly negatively correlated with neuroticism (--0.07). These results largely matched expectations, although the correlation with agreeableness was a bit larger than expected. These differences may be due in part to the context of the GFT study, wherein participants interacted with strangers (e.g., agreeable participants may be especially expressive in this context due to higher interpersonal trust and cooperation). The weak negative correlation with neuroticism suggests that this context may attenuate displays of negative emotion (e.g., to make a good first impression). Given this evidence of the latent variable's internal consistency and external validity, we felt confident using the factor scores as labels of emotional expressiveness in our predictive modeling experiments \cite{mcneish2020}.

\section{Measuring Behavior}
\label{sec:behavior}
In order to study the ways in which emotional expressiveness manifests in communicative behavior, we needed to measure participants' behavior. To do so, we designed measures of various visual and linguistic behavioral signals from the video recordings and transcripts. We based our choice of behavioral signals on measures that had been used successfully in personality perception and emotion recognition, as described in the literature reviewed in Section \ref{sec:eelit}. We decided not to include measures of acoustic behaviors (e.g., speech prosody) because of the high likelihood of imperfect audio source separation among the three participants in each group, particularly given the suboptimal recording quality and the frequency with which subjects spoke simultaneously. However, we do believe that acoustic behavioral signals are important to expressiveness and plan to explore them in future work after experimenting with different approaches for source separation.

\subsection{Visual Behavioral Signals}
For each participant's video, we used the OpenFace 2.0 \cite{baltrusaitis2018} toolkit to derive measures of visual behavioral signals. We began with frame-level estimates of the occurrence and intensity of action units (i.e., facial muscle movements) from the Facial Action Coding System \cite{ekman2002}, facial landmark (e.g., eye and mouth) coordinates, head pose (i.e., translation and rotation), and eye gaze angle. To reduce the effects of jitter, we averaged these measures over five-frame intervals, reducing the sampling rate from \SI{30}{\hertz} to \SI{6}{\hertz}. Furthermore, we used an affine transformation \cite{Hartley2004} to project each participant's facial landmark coordinates onto the average of all participant faces. This normalized the faces for size, translation, and rotation and isolated landmark movement from head movement. 

We then calculated the average number of action units present and the mean of all action units' intensities across all intervals. Our decision to aggregate all action units was motivated by a desire to capture overall facial movement without requiring OpenFace to reliably distinguish individual action units, which is a difficult task (especially when generalizing to novel datasets) \cite{cohn2019}. To measure the overall magnitude and speed of participants' facial and head motion, we computed the average displacement (i.e., distance traveled), velocity (i.e., the derivative of displacement), and acceleration (i.e., the derivative of velocity) for each facial landmark, head pose dimension, and eye gaze angle across five-frame intervals. Because measures for the individual facial landmark points were highly correlated, we summed across all landmark points. A full list of the 20 included visual behavioral signals can be found in \autoref{tab:features}.

\subsection{Linguistic Behavioral Signals}
For each participant's transcript, we used the LIWC2015 \cite{pennebaker2015} text analysis program to derive measures of their linguistic behavior across all utterances. LIWC2015 returns many measures that represent the percentage of the words in the text that fall into different categories (e.g., personal pronouns, common verbs, or swear words) as well as the output of several algorithms trained to predict linguistic dimensions (e.g., emotional tone, authenticity, and clout).

We selected categories and dimensions to include in our analyses based on several considerations. Because we were analyzing spoken words, we excluded measures that assume written text (e.g., punctuation and words per sentence). We prioritized categories and dimensions related to affective, social, and psychological processes (e.g., emotional tone, affiliation, and cognition). We also included categories related to parts of speech and language construction (e.g., pronouns, verbs, and fillers). 

In general, we chose to group linguistic behavioral signals into broader categories rather than including each individual category returned by LIWC2015 in order to reduce dimensionality and increase category reliability through aggregation. A full list of the 35 included linguistic behavioral signals can be found in \autoref{tab:features}.

\begin{table}[t]
    \caption{List of behavioral signals by modality}
    \label{tab:features}
    \footnotesize
    \begin{tabular}{l l l}
        \toprule
        Visual & \multicolumn{2}{l}{Linguistic} \\
        \midrule
        Mean Number of Action Units & Word Count & Cognitive Processes \\
        Mean Action Unit Intensity & Analytical Thinking & Sexual \\
        Mean Landmark Displacement & Clout & Ingestion \\
        Mean Landmark Velocity & Authentic & Affiliation \\
        Mean Landmark Acceleration & Emotional Tone & Reward \\
        Head Translation Displacement & Words > 6 Letters & Risk \\
        Head Translation Velocity & Personal Pronouns & Past Focus \\
        Head Translation Acceleration & Impersonal Pronouns & Present Focus \\
        Head Pitch Displacement & Common Adverbs & Future Focus \\
        Head Pitch Velocity & Negations & Work \\
        Head Pitch Acceleration & Common Verbs & Leisure \\
        Head Yaw Displacement & Common Adjectives & Home \\
        Head Yaw Velocity & Comparisons & Money \\
        Head Yaw Acceleration & Interrogatives & Swear words \\
        Head Roll Displacement & Quantifiers & Assent \\
        Head Roll Velocity & Positive Emotion & Nonfluencies \\
        Head Roll Acceleration & Negative Emotion & Fillers \\
        Gaze Angle Displacement & Social Processes &  \\
        Gaze Angle Velocity &  &  \\
        Gaze Angle Acceleration &  &  \\
        \bottomrule
    \end{tabular}
\end{table}

\section{Predictive Modeling}
We used predictive modeling techniques to explore our three research questions: (1) How well can we predict emotional expressiveness from multimodal representations of participants’ behavior? (2) How much do the different behavioral modalities contribute to the prediction of emotional expressiveness? (3) Which behavioral signals are reliably related to emotional expressiveness?

\subsection{Algorithms}
To investigate our first two questions, we needed predictive modeling algorithms that are well understood and well performing when trained on relatively small datasets. To investigate our third question, we needed at least one algorithm that could achieve competitive performance while providing interpretable results regarding the direction and strength of the relationship between each feature and label. Based on our literature review and prior experience, we decided to include the three algorithms described below.\footnote{Our goal in comparing multiple algorithms was not to determine which is best in general but rather to find one that performed well in this application. Proposing new algorithms is an interesting direction for future research but is beyond the scope of the current paper. Here, we focus on introducing a new topic area to the community and using existing algorithms to deepen scientific understanding of it.}

\subsubsection*{Elastic Net} 
This algorithm uses a mixture of L1 and L2 regularization on linear regression to reduce overfitting and multicollinearity, both of which are common problems when fitting linear models to data with many features \cite{zou2005}. Elastic Net models are fully interpretable, with the regression coefficients providing an indication of the direction and strength of the relationship between the features and the label. Our past work on visual expressiveness \cite{lin2019} found that Elastic Net's predictive performance exceeded that of several more complex models when trained on a small dataset.

\subsubsection*{Support Vector Regression (SVR)}
This algorithm transforms data into a higher-dimensional feature space and constructs a hyperplane that maximizes the number of points within some distance from it \cite{drucker1997}. Because Elastic Net models linear relationships only, we selected SVR with a radial basis function (RBF) kernel as a competing model to see whether modeling nonlinearities would yield better performance (at the cost of being able to interpret the relationships between features and labels). As described in Section \ref{sec:literature}, SVR and SVM have seen widespread use in the related fields of automatic personality perception and emotion recognition.

\subsubsection*{Multi-Layer Perceptron (MLP)}
Neural network models currently set the standard of performance for many relevant prediction tasks (as described in Section \ref{sec:literature}) due to their ability to capture complex nonlinear relationships (at the cost of interpretability, like SVR). Due to the relatively small size of our dataset and the absence of a temporal component, we constrained the complexity of our model to a simple MLP with one or two hidden layers. 

\subsection{Experimental Setup}
To explore our second research question about the relative importance of each modality, we fit three models per algorithm: one model using only visual behavioral signals as features, one model using only linguistic behavioral signals as features, and one multimodal model using both sets of behavioral signals as features. 

We performed 7-fold validation within 8-fold testing and repeated this process 20 times with different splits of the data. The folds were stratified to keep the proportion of labels in each quartile similar in each fold, and all three members of a given group were always assigned to the same fold (in order to prevent group-level dependencies from influencing performance estimates). Within each repetition, the folds were identical across algorithms and modalities to ensure that results were directly comparable. Root mean square error (RMSE) was used as the validation error measure.

We used the \textit{sklearn} implementations of these algorithms in our experiments and tuned hyperparameters via grid search. For Elastic Net, we tuned over $\alpha \in \{0.01, 0.05, 0.1, 0.5, 1.0, 1.1, 1.2,$ $1.3, 1.4, 1.5\}$ and $\lambda \in \{0.0, 0.1, 0.5, 0.7, 0.9, 0.95, 0.99,$ $1.0\}$, where $\alpha$ is the penalty term weight and $\lambda$ is the mixing parameter for L1 and L2 regularization. When $\lambda = 0.0$, Elastic Net becomes ridge regression, and when $\lambda = 1.0$, it becomes lasso regression \cite{zou2005}. For SVR, we tuned over $C \in \{2^{-5}, 2^{-4}, \dots,$ $2^{14}, 2^{15}\}$ and $\gamma \in \{2^{-15}, 2^{-14}, \dots, 2^2, 2^3\}$, where $C$ controls the strength of L2 regularization and $\gamma$ is the RBF kernel coefficient. For MLP regression, we tuned over the number of hidden layers (one or two), the number of hidden units per layer (64 or 128), and the L2 regularization term $\alpha \in \{0.0001, 0.001, 0.01, 0.05,$ $0.1, 0.5, 1.0, 1.1, 1.2, 1.3, 1.4,$ $1.5, 1.7, 1.9, 2.0, 3.0, 4.0, 5.0\}$.

To evaluate predictive performance during testing, we used three complementary performance metrics. First, we used RMSE as a measure of the distance between the predictions and labels in the same units as the labels. Given that the labels were standardized to have a standard deviation of 1, a perfect model would have $\text{RMSE}=0.0$, a model always predicting the mean would have $\text{RMSE}=1.0$, and a model with $\text{RMSE}>1.0$ would be worse than a model always predicting the mean. Second, we used the coefficient of determination $(R^2)$ as an estimate of the proportion of label variance explained by the features in a model. A perfect model would have $R^2=1.0$, a model always predicting the mean would have $R^2=0.0$, and a model with $R^2<0.0$ would be worse than a model always predicting the mean. Third, we used the Pearson correlation coefficient ($r$) as a standardized measure of the strength of the relationship between the predictions and labels. A perfect model would have $r=1.0$, an uninformative model would have $r=0.0$, and a model with $r<0.0$ would be misleading. Unlike the other two metrics, correlation does not require that the predictions and labels be identical---only that they be linearly related.

To compare the performance of models trained using different features, we used non-parametric bootstrap resampling with percentile confidence intervals and $p$-values \cite{efron1993}. Specifically, we created 2000 resamples ($n=160$) of the fold-level differences between models (through sampling with replacement) and, as in Table~\ref{tab:performance}, calculated the median as a robust measure of central tendency.

\subsection{Predictive Modeling Results}

\begin{table}[t]
    \caption{Performance of predictive models}
    \label{tab:performance}
    \sisetup{detect-all}
    \begin{tabular}{l l S[table-format=1.2] S[table-format=-1.2] S[table-format=1.2]}
    \toprule
    Algorithm & Modality & {RMSE} & {$R^2$} & {$r$} \\
    \midrule
    Elastic Net & Visual        & 0.89 & -0.03 & 0.45 \\
    Elastic Net & Linguistic    & 0.76 &  0.30 & 0.65 \\
    Elastic Net & Multimodal    & \bfseries 0.65 & \bfseries  0.45 & \bfseries 0.74 \\
    \midrule
    SVR & Visual                & 0.87 &  0.05 & 0.44 \\
    SVR & Linguistic            & 0.75 &  0.34 & 0.66 \\
    SVR & Multimodal            & 0.78 &  0.27 & 0.63 \\
    \midrule
    MLP & Visual                & 0.89 &  0.02 & 0.43 \\
    MLP & Linguistic            & 0.76 &  0.30 & 0.65 \\
    MLP & Multimodal            & 0.72 &  0.36 & 0.68 \\
    \bottomrule
    \end{tabular}
\end{table}


Because we had eight test folds in each of the 20 repetitions of our cross-validation procedure, our experiment yielded a distribution of 160 scores on each performance metric. In \autoref{tab:performance}, we present the median of each distribution as a robust measure of its central tendency. A full record of the results for all models 
is provided in the supplemental materials.

\begin{figure}[t]
    \centering
    \includegraphics[width=\linewidth]{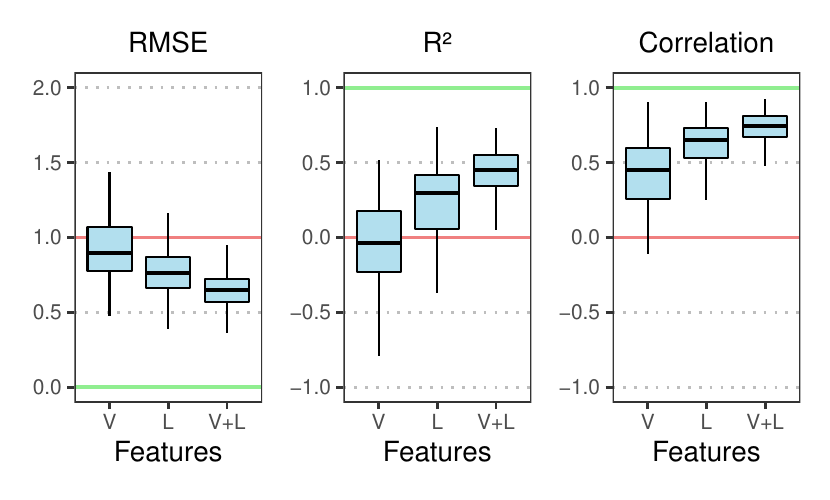}
    \caption{Performance of Elastic Net models by modality}
    \label{fig:en_all_performance}
    \Description{The multimodal models were better than the visual and linguistic models}
\end{figure}

\subsubsection*{Research Question 1}

To investigate our first research question about the prediction of emotional expressiveness from behavioral signals, we examined the best performing algorithm-modality combination: Elastic Net using multimodal behavioral signals as features. The predictions of this model were an average of 0.65 standard deviations from the corresponding label values, the features explained nearly half (45\%) of the variance in the labels, and the predictions were highly correlated with the labels (0.74). The strength of the correlation in particular suggests that the predictions of this model could be used as a proxy for human ratings in settings where preserving linear relationships is sufficient. However, the fact that more than half of the variance in the labels was unexplained by the features suggests that there is still more to emotional expressiveness than what our behavioral measures were able to capture.

We suspect that much of the unexplained variance is related to acoustic behavioral signals (e.g., speech prosody); the rest may be related to visual and linguistic signals that our representations did not capture (e.g., specific facial actions or gestures) or cultural and individual differences in patterns of expressive behavior. 

\subsubsection*{Research Question 2}

\begin{table}[t]
    \caption{Comparison of modalities for Elastic Net}
    \label{tab:modalities}
    \begin{tabular}{
        l 
        S[table-format = -1.2]
        >{{[}} 
        S[table-format = -1.2,table-space-text-pre={[}] 
        @{,\,} 
        S[table-format = -1.2,table-space-text-post={]}] 
        <{{]}} 
        S[table-format=0.3, add-integer-zero=false, table-comparator=true]
    }
    \toprule
    Difference in RMSE & {$\Delta$} & \multicolumn{2}{c}{95\% CI} & {$p$} \\ 
    \midrule
    Visual -- Linguistic & 0.16 & 0.11 & 0.18 & <.001 \\ 
    Visual -- Multimodal & 0.25 & 0.22 & 0.27 & <.001 \\ 
    Linguistic -- Multimodal & 0.10 & 0.08 & 0.11 & <.001 \\ 
    \midrule
    Difference in $R^2$ & {$\Delta$} & \multicolumn{2}{c}{95\% CI} & {$p$} \\ 
    \midrule
    Visual -- Linguistic & -0.31 & -0.36 & -0.19 & <.001 \\ 
    Visual -- Multimodal & -0.46 & -0.53 & -0.41 & <.001 \\ 
    Linguistic -- Multimodal & -0.17 & -0.21 & -0.15 & <.001 \\ 
    \midrule
    Difference in $r$ & {$\Delta$} & \multicolumn{2}{c}{95\% CI} & {$p$} \\ 
    \midrule
    Visual -- Linguistic & -0.21 & -0.27 & -0.14 & <.001 \\ 
    Visual -- Multimodal & -0.30 & -0.34 & -0.25 & <.001 \\ 
    Linguistic -- Multimodal & -0.11 & -0.13 & -0.09 & <.001 \\ 
    \bottomrule
    \end{tabular}
\end{table}

\begin{figure}[t]
    \centering
    \includegraphics[width=\linewidth]{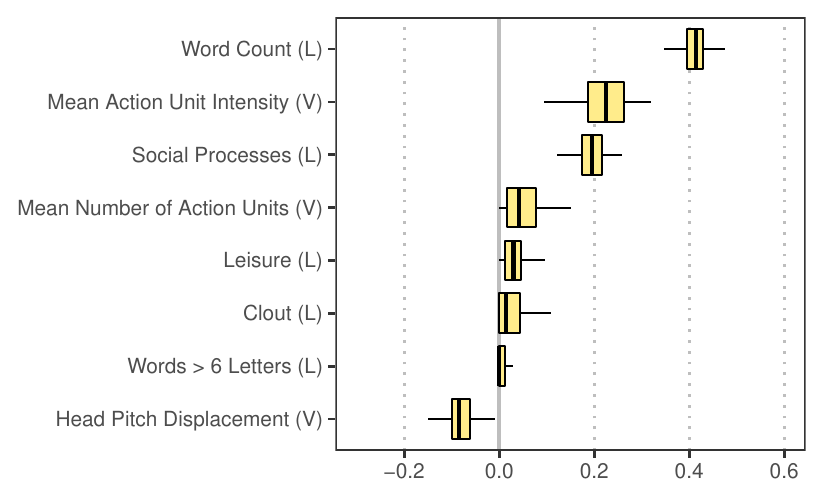}
    \caption{Coefficients from multimodal Elastic Net models}
    \label{fig:en_vl_coeff}
\end{figure}

To investigate our second research question about the relative importance of the visual and linguistic modalities (given our selected behavioral representations), we compared the performance of the three Elastic Net models. The results of the modality comparison are provided in Table~\ref{tab:modalities}.

Our first finding was that the linguistic model significantly outperformed the visual model on all performance metrics. This result may suggest that emotional expressiveness is conveyed more through linguistic behavioral signals than through visual behavioral signals. Alternatively, it may suggest that our linguistic representations (based on manual transcripts) were more reliable than our visual representations (based on automatic tracking).

Our second finding was that the multimodal model significantly outperformed the the visual model and the linguistic model on all performance metrics. These results suggest that both modalities made unique contributions to the prediction and that both are worth including in future work on emotional expressiveness.

\subsubsection*{Research Question 3}

To investigate our third research question about the relationships between specific behavioral signals and emotional expressiveness, we examined the distributions of the regression coefficients across all tuned multimodal Elastic Net models. \autoref{fig:en_vl_coeff} depicts the distributions with a nonzero median (and all distributions are depicted in the supplemental materials). 

With a median coefficient of around $0.4$, the total number of words spoken (\textit{Word Count}) was the behavioral signal with the strongest relationship to emotional expressiveness. As language is a primary channel of human communication, it is not surprising that participants rated higher on emotional expressiveness tended to speak more. That is, speaking more likely gave participants additional opportunities to express themselves. 

Two other behavioral signals had median coefficients around $0.2$: the average magnitude of all facial movements \textit{(Mean Action Unit Intensity)} and the proportion of words that fell in the \textit{Social Processes} category (e.g., family and friends). In general, more intense facial movements correspond to stronger emotion \cite{keltner2003}, and since we define emotional expressiveness as the tendency to outwardly display emotion, it follows that participants who show stronger emotion are more expressive. This result is also consistent with our prior findings on the role of action unit intensity in momentary emotional expressiveness \cite{lin2019}. The relationship between the \emph{Social Processes} linguistic signal and emotional expressiveness suggests a link between sociability and emotional expressiveness that is consistent with the correlation between emotional expressiveness and self-reported extraversion observed in this study and others \cite{gross1995,kring1994}. Specifically, individuals who socialize more (or who speak more about their social habits and activities) may tend to be more expressive not only of such activities but also of their emotions.

Finally, five other behavioral signals had median coefficients between $0.0$ and $0.1$ or between $0.0$ and $-0.1$. Participants rated higher on emotional expressiveness had more facial movements (\textit{Mean Number of Action Units}), more words that fell in the \textit{Leisure} category (e.g., cook, chat, movie), higher scores on the \textit{Clout} dimension (i.e., speaking with confidence and authority), more words longer than six letters (\textit{Words $>$ 6 Letters}), and smaller vertical head motions (\textit{Head Pitch Displacement}). Some of the relationships between these signals (i.e., number of action units and leisure words) are fairly similar in their interpretation to the relationships described in the previous paragraph. We also believe that the last finding---the negative coefficient for vertical head motion---may reflect that highly expressive participants spend more time talking and less time listening (as nodding is often a sign of listening \cite{pasupathi1999responsive}). The associations between emotional expressiveness and \emph{Clout} and long words are more novel. These results point to possible positive relationships between emotional expressiveness and assertiveness (a facet of extraversion). Individuals who exhibit greater confidence in word choice, whether through more dominant or more advanced vocabulary, may tend to be more emotionally expressive.

\section{Discussion}
In this paper, we added highly reliable transcripts of spoken words and human perceptual ratings of emotional expressiveness to an existing video database of unstructured social interactions. We proposed and tested a series of models encompassing traditional statistical modeling and machine learning to predict emotional expressiveness from visual and linguistic features. Encouragingly, we found that the multimodal statistical model (Elastic Net) performed reasonably well, explaining nearly half of the variance in the perceptual ratings. We then examined what this model learned in predicting emotional expressiveness from behavioral signals. Our results uncovered novel relationships between emotional expressiveness and behaviors related to traits like sociability, assertiveness, and confidence. Our most important visual features---those derived from action unit measures---converge with prior findings on the prediction and characterization of facial expressiveness in a very different context \cite{lin2019}. The consistency of these results suggests a persistent role of facial action units in signaling emotional expressiveness.


Our results should be considered in light of several points. First, the GFT dataset (like many others in affective computing) included only healthy young adults from a single geographic area, and it is unknown how well our findings would generalize to other populations. Second, the context of the participants' behavior (i.e., the group formation task) was unstructured and more naturalistic than previous work, but the interactions still took place in a laboratory setting. Finally, the relatively small size of the dataset (in terms of both the number of participants and the amount of time each participant was observed) likely limited the performance our predictive models were able to achieve. These limitations can all be addressed through more research in additional populations and contexts, and we hope this paper will serve to encourage other interdisciplinary researchers to study emotional expressiveness.

In conclusion, we laid the foundation for the computational study of \textit{emotional expressiveness}, which captures to the degree to which a person tends to outwardly display their emotions. We developed a method to collect human ratings of emotional expressiveness with high reliability and used latent variable modeling techniques to refine these ratings. We then constructed models to predict and characterize these ratings in terms of interpretable visual and linguistic behavioral signals. This line of research holds the potential to improve emotional inferences through personalized (i.e., expressiveness-aware) algorithms, develop more satisfying and efficient robots and virtual agents, and aid medical professionals in the assessment of psychiatric and neurological conditions.

\section*{Acknowledgements}
This paper is based upon work supported by the National Science Foundation (Awards 1722822 and 1734868) and National Institutes of Health (Awards R01-AA015773, R01-MH096951, U01-MH116925, and U01-MH116923). Any opinions and recommendations expressed in this material are those of the authors alone.

\section*{Supplemental Materials}
All source code from this paper can be found at \url{https://osf.io/umzar/}. The Sayette GFT database \cite{girard2017a} is freely available for academic purposes (to request access, view the wiki at \url{https://osf.io/7wcyz/}).

\bibliographystyle{ACM-Reference-Format}
\bibliography{ref}


\begin{thebibliography}{68}


\ifx \showCODEN    \undefined \def \showCODEN     #1{\unskip}     \fi
\ifx \showDOI      \undefined \def \showDOI       #1{#1}\fi
\ifx \showISBNx    \undefined \def \showISBNx     #1{\unskip}     \fi
\ifx \showISBNxiii \undefined \def \showISBNxiii  #1{\unskip}     \fi
\ifx \showISSN     \undefined \def \showISSN      #1{\unskip}     \fi
\ifx \showLCCN     \undefined \def \showLCCN      #1{\unskip}     \fi
\ifx \shownote     \undefined \def \shownote      #1{#1}          \fi
\ifx \showarticletitle \undefined \def \showarticletitle #1{#1}   \fi
\ifx \showURL      \undefined \def \showURL       {\relax}        \fi
\providecommand\bibfield[2]{#2}
\providecommand\bibinfo[2]{#2}
\providecommand\natexlab[1]{#1}
\providecommand\showeprint[2][]{arXiv:#2}

\bibitem[\protect\citeauthoryear{{American Psychiatric Association}}{{American
  Psychiatric Association}}{2013}]%
        {americanpsychiatricassociation2013}
\bibfield{author}{\bibinfo{person}{{American Psychiatric Association}}.}
  \bibinfo{year}{2013}\natexlab{}.
\newblock \bibinfo{booktitle}{\emph{Diagnostic and Statistical Manual of Mental
  Disorders} (\bibinfo{edition}{fifth} ed.)}.
\newblock \bibinfo{publisher}{{}American Psychiatric Association},
  \bibinfo{address}{{Washington, DC}}.
\newblock


\bibitem[\protect\citeauthoryear{Arslan, Walther, and Tata}{Arslan
  et~al\mbox{.}}{2020}]%
        {arslan2020}
\bibfield{author}{\bibinfo{person}{Ruben~C. Arslan},
  \bibinfo{person}{Matthias~P. Walther}, {and} \bibinfo{person}{Cyril~S.
  Tata}.} \bibinfo{year}{2020}\natexlab{}.
\newblock \showarticletitle{Formr: {{A}} Study Framework Allowing for Automated
  Feedback Generation and Complex Longitudinal Experience-Sampling Studies
  Using {{R}}}.
\newblock \bibinfo{journal}{\emph{Behavior Research Methods}}
  \bibinfo{volume}{52} (\bibinfo{year}{2020}), \bibinfo{pages}{376--387}.
\newblock


\bibitem[\protect\citeauthoryear{Baltrusaitis, Zadeh, Lim, and
  Morency}{Baltrusaitis et~al\mbox{.}}{2018}]%
        {baltrusaitis2018}
\bibfield{author}{\bibinfo{person}{Tadas Baltrusaitis}, \bibinfo{person}{Amir
  Zadeh}, \bibinfo{person}{Yao~Chong Lim}, {and}
  \bibinfo{person}{Louis-Philippe Morency}.} \bibinfo{year}{2018}\natexlab{}.
\newblock \showarticletitle{{{OpenFace}} 2.0: {{Facial}} Behavior Analysis
  Toolkit}. In \bibinfo{booktitle}{\emph{Proceedings of the 13th {{IEEE
  International Conference}} on {{Automatic Face}} and {{Gesture
  Recognition}}}}. \bibinfo{publisher}{{IEEE}}, \bibinfo{address}{{Xi'an,
  China}}, \bibinfo{pages}{59--66}.
\newblock


\bibitem[\protect\citeauthoryear{Barrett, Adolphs, Marsella, Martinez, and
  Pollak}{Barrett et~al\mbox{.}}{2019}]%
        {barrett2019}
\bibfield{author}{\bibinfo{person}{Lisa~Feldman Barrett},
  \bibinfo{person}{Ralph Adolphs}, \bibinfo{person}{Stacy Marsella},
  \bibinfo{person}{Aleix~M. Martinez}, {and} \bibinfo{person}{Seth~D. Pollak}.}
  \bibinfo{year}{2019}\natexlab{}.
\newblock \showarticletitle{Emotional Expressions Reconsidered: {{Challenges}}
  to Inferring Emotion from Human Facial Movements}.
\newblock \bibinfo{journal}{\emph{Psychological Science in the Public
  Interest}} \bibinfo{volume}{20}, \bibinfo{number}{1} (\bibinfo{year}{2019}),
  \bibinfo{pages}{1--68}.
\newblock


\bibitem[\protect\citeauthoryear{Biel and Gatica-Perez}{Biel and
  Gatica-Perez}{2012}]%
        {biel2012youtube}
\bibfield{author}{\bibinfo{person}{Joan-Isaac Biel} {and}
  \bibinfo{person}{Daniel Gatica-Perez}.} \bibinfo{year}{2012}\natexlab{}.
\newblock \showarticletitle{The youtube lens: Crowdsourced personality
  impressions and audiovisual analysis of vlogs}.
\newblock \bibinfo{journal}{\emph{IEEE Transactions on Multimedia}}
  \bibinfo{volume}{15}, \bibinfo{number}{1} (\bibinfo{year}{2012}),
  \bibinfo{pages}{41--55}.
\newblock


\bibitem[\protect\citeauthoryear{Chastagnol and Devillers}{Chastagnol and
  Devillers}{2012}]%
        {chastagnol2012personality}
\bibfield{author}{\bibinfo{person}{Cl{\'e}ment Chastagnol} {and}
  \bibinfo{person}{Laurence Devillers}.} \bibinfo{year}{2012}\natexlab{}.
\newblock \showarticletitle{Personality Traits Detection using a Parallelized
  Modified SFFS Algorithm}. In \bibinfo{booktitle}{\emph{Proceedings of the
  13th Annual Conference of the International Speech Communication
  Association}}. \bibinfo{publisher}{ISCA}, \bibinfo{address}{Portland, OR},
  \bibinfo{pages}{266--269}.
\newblock


\bibitem[\protect\citeauthoryear{Cohen, Sebe, Garg, Chen, and Huang}{Cohen
  et~al\mbox{.}}{2003}]%
        {cohen2003facial}
\bibfield{author}{\bibinfo{person}{Ira Cohen}, \bibinfo{person}{Nicu Sebe},
  \bibinfo{person}{Ashutosh Garg}, \bibinfo{person}{Lawrence~S. Chen}, {and}
  \bibinfo{person}{Thomas~S. Huang}.} \bibinfo{year}{2003}\natexlab{}.
\newblock \showarticletitle{Facial Expression Recognition from Video Sequences:
  Temporal and Static Modeling}.
\newblock \bibinfo{journal}{\emph{Computer Vision and Image Understanding}}
  \bibinfo{volume}{91}, \bibinfo{number}{1-2} (\bibinfo{year}{2003}),
  \bibinfo{pages}{160--187}.
\newblock


\bibitem[\protect\citeauthoryear{Cohn, Ertugrul, Chu, Girard, and Hammal}{Cohn
  et~al\mbox{.}}{2019}]%
        {cohn2019}
\bibfield{author}{\bibinfo{person}{Jeffrey~F. Cohn}, \bibinfo{person}{Itir~Onal
  Ertugrul}, \bibinfo{person}{Wen-Sheng Chu}, \bibinfo{person}{Jeffrey~M.
  Girard}, {and} \bibinfo{person}{Zakia Hammal}.}
  \bibinfo{year}{2019}\natexlab{}.
\newblock \showarticletitle{Affective Facial Computing: {{Generalizability}}
  across Domains}.
\newblock In \bibinfo{booktitle}{\emph{Multimodal Behavior Analysis in the
  Wild: {{Advances}} and Challenges}},
  \bibfield{editor}{\bibinfo{person}{Xavier {Alameda-Pineda}},
  \bibinfo{person}{Elisa Ricci}, {and} \bibinfo{person}{Nicu Sebe}} (Eds.).
  \bibinfo{publisher}{{Academic Press}}, \bibinfo{address}{Cambridge, MA},
  \bibinfo{pages}{407--441}.
\newblock
\showISBNx{978-0-12-814601-9}


\bibitem[\protect\citeauthoryear{Costa and McCrae}{Costa and McCrae}{1992}]%
        {costa1992}
\bibfield{author}{\bibinfo{person}{Paul~T. Costa} {and}
  \bibinfo{person}{Robert~R. McCrae}.} \bibinfo{year}{1992}\natexlab{}.
\newblock \bibinfo{booktitle}{\emph{Revised {{NEO Personality Inventory}}
  ({{NEO PI}}-{{R}}) and {{NEO Five}}-{{Factor Inventory}} ({{NEO}}-{{FFI}})}}.
\newblock \bibinfo{publisher}{{Psychological Assessment Resources}},
  \bibinfo{address}{{Odessa, FL}}.
\newblock


\bibitem[\protect\citeauthoryear{Cowen, Sauter, Tracy, and Keltner}{Cowen
  et~al\mbox{.}}{2019}]%
        {cowen2019}
\bibfield{author}{\bibinfo{person}{Alan Cowen}, \bibinfo{person}{Disa Sauter},
  \bibinfo{person}{Jessica~L. Tracy}, {and} \bibinfo{person}{Dacher Keltner}.}
  \bibinfo{year}{2019}\natexlab{}.
\newblock \showarticletitle{Mapping the Passions: {{Toward}} a High-Dimensional
  Taxonomy of Emotional Experience and Expression}.
\newblock \bibinfo{journal}{\emph{Psychological Science in the Public
  Interest}} \bibinfo{volume}{20}, \bibinfo{number}{1} (\bibinfo{year}{2019}),
  \bibinfo{pages}{69--90}.
\newblock


\bibitem[\protect\citeauthoryear{Cred{\'e}, Harms, Niehorster, and
  {Gaye-Valentine}}{Cred{\'e} et~al\mbox{.}}{2012}]%
        {crede2012}
\bibfield{author}{\bibinfo{person}{Marcus Cred{\'e}}, \bibinfo{person}{Peter
  Harms}, \bibinfo{person}{Sarah Niehorster}, {and} \bibinfo{person}{Andrea
  {Gaye-Valentine}}.} \bibinfo{year}{2012}\natexlab{}.
\newblock \showarticletitle{An Evaluation of the Consequences of Using Short
  Measures of the {{Big Five}} Personality Traits}.
\newblock \bibinfo{journal}{\emph{Journal of Personality and Social
  Psychology}} \bibinfo{volume}{102}, \bibinfo{number}{4}
  (\bibinfo{year}{2012}), \bibinfo{pages}{874--888}.
\newblock
\showISSN{0022-3514}


\bibitem[\protect\citeauthoryear{DeCoster, Iselin, and Gallucci}{DeCoster
  et~al\mbox{.}}{2009}]%
        {decoster2009}
\bibfield{author}{\bibinfo{person}{Jamie DeCoster},
  \bibinfo{person}{Anne-Marie~R. Iselin}, {and} \bibinfo{person}{Marcello
  Gallucci}.} \bibinfo{year}{2009}\natexlab{}.
\newblock \showarticletitle{A Conceptual and Empirical Examination of
  Justifications for Dichotomization}.
\newblock \bibinfo{journal}{\emph{Psychological Methods}} \bibinfo{volume}{14},
  \bibinfo{number}{4} (\bibinfo{year}{2009}), \bibinfo{pages}{349--366}.
\newblock
\showISSN{1082-989X}


\bibitem[\protect\citeauthoryear{Dhall, Goecke, Joshi, Wagner, and
  Gedeon}{Dhall et~al\mbox{.}}{2013}]%
        {dhall2013emotion}
\bibfield{author}{\bibinfo{person}{Abhinav Dhall}, \bibinfo{person}{Roland
  Goecke}, \bibinfo{person}{Jyoti Joshi}, \bibinfo{person}{Michael Wagner},
  {and} \bibinfo{person}{Tom Gedeon}.} \bibinfo{year}{2013}\natexlab{}.
\newblock \showarticletitle{Emotion Recognition in the Wild Challenge 2013}. In
  \bibinfo{booktitle}{\emph{Proceedings of the 15th ACM on International
  Conference on Multimodal Interaction}}. \bibinfo{publisher}{ACM},
  \bibinfo{address}{Sydney, Australia}, \bibinfo{pages}{509--516}.
\newblock


\bibitem[\protect\citeauthoryear{Dhall, Kaur, Goecke, and Gedeon}{Dhall
  et~al\mbox{.}}{2018}]%
        {dhall2018emotiw}
\bibfield{author}{\bibinfo{person}{Abhinav Dhall}, \bibinfo{person}{Amanjot
  Kaur}, \bibinfo{person}{Roland Goecke}, {and} \bibinfo{person}{Tom Gedeon}.}
  \bibinfo{year}{2018}\natexlab{}.
\newblock \showarticletitle{EmotiW 2018: Audio-Video, Student Engagement and
  Group-level Affect Prediction}. In \bibinfo{booktitle}{\emph{Proceedings of
  the 20th ACM International Conference on Multimodal Interaction}}.
  \bibinfo{publisher}{ACM}, \bibinfo{address}{Boulder, CO},
  \bibinfo{pages}{653--656}.
\newblock


\bibitem[\protect\citeauthoryear{Dhall, Ramana~Murthy, Goecke, Joshi, and
  Gedeon}{Dhall et~al\mbox{.}}{2015}]%
        {dhall2015video}
\bibfield{author}{\bibinfo{person}{Abhinav Dhall}, \bibinfo{person}{O.~V.
  Ramana~Murthy}, \bibinfo{person}{Roland Goecke}, \bibinfo{person}{Jyoti
  Joshi}, {and} \bibinfo{person}{Tom Gedeon}.} \bibinfo{year}{2015}\natexlab{}.
\newblock \showarticletitle{Video and image based emotion recognition
  challenges in the wild: Emotiw 2015}. In
  \bibinfo{booktitle}{\emph{Proceedings of the 17th ACM International
  Conference on Multimodal Interaction}}. \bibinfo{publisher}{ACM},
  \bibinfo{address}{Seattle, WA}, \bibinfo{pages}{423--426}.
\newblock


\bibitem[\protect\citeauthoryear{Donahue, Jia, Vinyals, Hoffman, Zhang, Tzeng,
  and Darrell}{Donahue et~al\mbox{.}}{2014}]%
        {donahue2014decaf}
\bibfield{author}{\bibinfo{person}{Jeff Donahue}, \bibinfo{person}{Yangqing
  Jia}, \bibinfo{person}{Oriol Vinyals}, \bibinfo{person}{Judy Hoffman},
  \bibinfo{person}{Ning Zhang}, \bibinfo{person}{Eric Tzeng}, {and}
  \bibinfo{person}{Trevor Darrell}.} \bibinfo{year}{2014}\natexlab{}.
\newblock \showarticletitle{Decaf: A deep convolutional activation feature for
  generic visual recognition}. In \bibinfo{booktitle}{\emph{Proceedings of the
  31st International Conference on Machine Learning}}.
  \bibinfo{publisher}{IMLS}, \bibinfo{address}{Beijing, China},
  \bibinfo{pages}{647--655}.
\newblock


\bibitem[\protect\citeauthoryear{Drucker, Burges, Kaufman, Smola, and
  Vapnik}{Drucker et~al\mbox{.}}{1997}]%
        {drucker1997}
\bibfield{author}{\bibinfo{person}{Harris Drucker},
  \bibinfo{person}{Christopher J.~C. Burges}, \bibinfo{person}{Linda Kaufman},
  \bibinfo{person}{Alex~J. Smola}, {and} \bibinfo{person}{Vladimir Vapnik}.}
  \bibinfo{year}{1997}\natexlab{}.
\newblock \showarticletitle{Support Vector Regression Machines}. In
  \bibinfo{booktitle}{\emph{Advances in {{Neural Information Processing
  Systems}}}}, Vol.~\bibinfo{volume}{9}. \bibinfo{publisher}{{MIT Press}},
  \bibinfo{address}{{Cambridge, MA}}, \bibinfo{pages}{155--161}.
\newblock


\bibitem[\protect\citeauthoryear{Efron and Tibshirani}{Efron and
  Tibshirani}{1993}]%
        {efron1993}
\bibfield{author}{\bibinfo{person}{Bradley Efron} {and}
  \bibinfo{person}{Robert~J. Tibshirani}.} \bibinfo{year}{1993}\natexlab{}.
\newblock \bibinfo{booktitle}{\emph{An Introduction to the Bootstrap}}.
\newblock \bibinfo{publisher}{{Chapman and Hall}}, \bibinfo{address}{{New York,
  NY}}.
\newblock


\bibitem[\protect\citeauthoryear{Ekman, Friesen, and Hager}{Ekman
  et~al\mbox{.}}{2002}]%
        {ekman2002}
\bibfield{author}{\bibinfo{person}{Paul Ekman}, \bibinfo{person}{Wallace~V.
  Friesen}, {and} \bibinfo{person}{John Hager}.}
  \bibinfo{year}{2002}\natexlab{}.
\newblock \bibinfo{booktitle}{\emph{Facial Action Coding System: {{A}}
  Technique for the Measurement of Facial Movement}}.
\newblock \bibinfo{publisher}{{Research Nexus}}, \bibinfo{address}{{Salt Lake
  City, UT}}.
\newblock


\bibitem[\protect\citeauthoryear{Flake, Pek, and Hehman}{Flake
  et~al\mbox{.}}{2017}]%
        {flake2017}
\bibfield{author}{\bibinfo{person}{Jessica~K. Flake}, \bibinfo{person}{Jolynn
  Pek}, {and} \bibinfo{person}{Eric Hehman}.} \bibinfo{year}{2017}\natexlab{}.
\newblock \showarticletitle{Construct Validation in Social and Personality
  Research: {{Current}} Practice and Recommendations}.
\newblock \bibinfo{journal}{\emph{Social Psychological and Personality
  Science}} \bibinfo{volume}{8}, \bibinfo{number}{4} (\bibinfo{year}{2017}),
  \bibinfo{pages}{370--378}.
\newblock


\bibitem[\protect\citeauthoryear{Friedman, Prince, and Riggio}{Friedman
  et~al\mbox{.}}{1980}]%
        {friedman1980}
\bibfield{author}{\bibinfo{person}{Howard~S. Friedman},
  \bibinfo{person}{Louise~M. Prince}, {and} \bibinfo{person}{Ronald~E.
  Riggio}.} \bibinfo{year}{1980}\natexlab{}.
\newblock \showarticletitle{Understanding and Assessing Nonverbal
  Expressiveness: {{The Affective Communication Test}}}.
\newblock \bibinfo{journal}{\emph{Journal of Personality and Social
  Psychology}} \bibinfo{volume}{39}, \bibinfo{number}{2}
  (\bibinfo{year}{1980}), \bibinfo{pages}{333--351}.
\newblock


\bibitem[\protect\citeauthoryear{{Garnier-Villarreal} and
  Jorgensen}{{Garnier-Villarreal} and Jorgensen}{2020}]%
        {garnier-villarreal2020}
\bibfield{author}{\bibinfo{person}{Mauricio {Garnier-Villarreal}} {and}
  \bibinfo{person}{Terrence~D. Jorgensen}.} \bibinfo{year}{2020}\natexlab{}.
\newblock \showarticletitle{Adapting Fit Indices for {{Bayesian}} Structural
  Equation Modeling: {{Comparison}} to Maximum Likelihood}.
\newblock \bibinfo{journal}{\emph{Psychological Methods}} \bibinfo{volume}{25},
  \bibinfo{number}{1} (\bibinfo{year}{2020}), \bibinfo{pages}{46--70}.
\newblock


\bibitem[\protect\citeauthoryear{Girard}{Girard}{2020}]%
        {girard2020}
\bibfield{author}{\bibinfo{person}{Jeffrey~M. Girard}.}
  \bibinfo{year}{2020}\natexlab{}.
\newblock \bibinfo{title}{agreement: {{An R}} Package for the Tidy Analysis of
  Agreement and Reliability}.
\newblock
\newblock
\urldef\tempurl%
\url{https://github.com/jmgirard/agreement}
\showURL{%
\tempurl}


\bibitem[\protect\citeauthoryear{Girard, Chu, Jeni, Cohn, De~La~Torre, and
  Sayette}{Girard et~al\mbox{.}}{2017}]%
        {girard2017a}
\bibfield{author}{\bibinfo{person}{Jeffrey~M. Girard},
  \bibinfo{person}{Wen-Sheng Chu}, \bibinfo{person}{Laszlo~A. Jeni},
  \bibinfo{person}{Jeffrey~F. Cohn}, \bibinfo{person}{F. De~La~Torre}, {and}
  \bibinfo{person}{Michael~A. Sayette}.} \bibinfo{year}{2017}\natexlab{}.
\newblock \showarticletitle{Sayette {{Group Formation Task}} ({{GFT}})
  {{Spontaneous Facial Expression Database}}}. In
  \bibinfo{booktitle}{\emph{Proceedings of the 12th {{IEEE International
  Conference}} on {{Automatic Face}} and {{Gesture Recognition}}}}.
  \bibinfo{publisher}{IEEE}, \bibinfo{address}{Washington, DC},
  \bibinfo{pages}{581--588}.
\newblock
\showISBNx{978-1-5090-4023-0}


\bibitem[\protect\citeauthoryear{Gross and John}{Gross and John}{1995}]%
        {gross1995}
\bibfield{author}{\bibinfo{person}{James~J. Gross} {and}
  \bibinfo{person}{Oliver~P. John}.} \bibinfo{year}{1995}\natexlab{}.
\newblock \showarticletitle{Facets of Emotional Expressivity: {{Three}}
  Self-Report Factors and Their Correlates}.
\newblock \bibinfo{journal}{\emph{Personality and Individual Differences}}
  \bibinfo{volume}{19}, \bibinfo{number}{4} (\bibinfo{year}{1995}),
  \bibinfo{pages}{555--568}.
\newblock
\showISSN{01918869}


\bibitem[\protect\citeauthoryear{Gunes and Schuller}{Gunes and
  Schuller}{2013}]%
        {gunes2013a}
\bibfield{author}{\bibinfo{person}{Hatice Gunes} {and}
  \bibinfo{person}{Bj{\"o}rn~W. Schuller}.} \bibinfo{year}{2013}\natexlab{}.
\newblock \showarticletitle{Categorical and Dimensional Affect Analysis in
  Continuous Input: {{Current}} Trends and Future Directions}.
\newblock \bibinfo{journal}{\emph{Image and Vision Computing}}
  \bibinfo{volume}{31}, \bibinfo{number}{2} (\bibinfo{year}{2013}),
  \bibinfo{pages}{120--136}.
\newblock


\bibitem[\protect\citeauthoryear{Gwet}{Gwet}{2014}]%
        {gwet2014}
\bibfield{author}{\bibinfo{person}{Kilem~L. Gwet}.}
  \bibinfo{year}{2014}\natexlab{}.
\newblock \bibinfo{booktitle}{\emph{Handbook of Inter-Rater Reliability:
  {{The}} Definitive Guide to Measuring the Extent of Agreement among Raters}
  (\bibinfo{edition}{fourth} ed.)}.
\newblock \bibinfo{publisher}{{Advanced Analytics}},
  \bibinfo{address}{{Gaithersburg, MD}}.
\newblock


\bibitem[\protect\citeauthoryear{Hamacher, Bianchi-Berthouze, Pipe, and
  Eder}{Hamacher et~al\mbox{.}}{2016}]%
        {hamacher2016believing}
\bibfield{author}{\bibinfo{person}{Adriana Hamacher}, \bibinfo{person}{Nadia
  Bianchi-Berthouze}, \bibinfo{person}{Anthony~G. Pipe}, {and}
  \bibinfo{person}{Kerstin Eder}.} \bibinfo{year}{2016}\natexlab{}.
\newblock \showarticletitle{Believing in BERT: Using Expressive Communication
  to Enhance Trust and Counteract Operational Error in Physical Human-Robot
  Interaction}. In \bibinfo{booktitle}{\emph{Proceedings of the 25th IEEE
  International Symposium on Robot and Human Interactive Communication}}.
  \bibinfo{publisher}{IEEE}, \bibinfo{address}{New York, NY},
  \bibinfo{pages}{493--500}.
\newblock


\bibitem[\protect\citeauthoryear{Hartley and Zisserman}{Hartley and
  Zisserman}{2004}]%
        {Hartley2004}
\bibfield{author}{\bibinfo{person}{Richard~I. Hartley} {and}
  \bibinfo{person}{Andrew Zisserman}.} \bibinfo{year}{2004}\natexlab{}.
\newblock \bibinfo{booktitle}{\emph{Multiple View Geometry in Computer Vision}
  (\bibinfo{edition}{second} ed.)}.
\newblock \bibinfo{publisher}{Cambridge University Press},
  \bibinfo{address}{Cambridge, England}.
\newblock


\bibitem[\protect\citeauthoryear{Hu and Bentler}{Hu and Bentler}{1999}]%
        {hu1999}
\bibfield{author}{\bibinfo{person}{Li-tze Hu} {and} \bibinfo{person}{Peter~M.
  Bentler}.} \bibinfo{year}{1999}\natexlab{}.
\newblock \showarticletitle{Cutoff Criteria for Fit Indexes in Covariance
  Structure Analysis: {{Conventional}} Criteria versus New Alternatives}.
\newblock \bibinfo{journal}{\emph{Structural Equation Modeling}}
  \bibinfo{volume}{6} (\bibinfo{year}{1999}), \bibinfo{pages}{1--55}.
\newblock


\bibitem[\protect\citeauthoryear{Kapoor and Picard}{Kapoor and Picard}{2005}]%
        {kapoor2005multimodal}
\bibfield{author}{\bibinfo{person}{Ashish Kapoor} {and}
  \bibinfo{person}{Rosalind~W. Picard}.} \bibinfo{year}{2005}\natexlab{}.
\newblock \showarticletitle{Multimodal Affect Recognition in Learning
  Environments}. In \bibinfo{booktitle}{\emph{Proceedings of the 13th ACM
  International Conference on Multimedia}}. \bibinfo{publisher}{ACM},
  \bibinfo{address}{Singapore}, \bibinfo{pages}{677--682}.
\newblock


\bibitem[\protect\citeauthoryear{Keltner, Ekman, Gonzaga, and Beer}{Keltner
  et~al\mbox{.}}{2003}]%
        {keltner2003}
\bibfield{author}{\bibinfo{person}{Dacher Keltner}, \bibinfo{person}{Paul
  Ekman}, \bibinfo{person}{Gian~C. Gonzaga}, {and} \bibinfo{person}{Jennifer
  Beer}.} \bibinfo{year}{2003}\natexlab{}.
\newblock \showarticletitle{Facial Expression of Emotion}.
\newblock In \bibinfo{booktitle}{\emph{Handbook of Affective Sciences}},
  \bibfield{editor}{\bibinfo{person}{H.~Hill Goldsmith},
  \bibinfo{person}{Klaus~R. Scherer}, {and} \bibinfo{person}{Richard~J.
  Davidson}} (Eds.). \bibinfo{publisher}{{Oxford University Press}},
  \bibinfo{address}{Oxford, England}, \bibinfo{pages}{415--432}.
\newblock


\bibitem[\protect\citeauthoryear{King and Emmons}{King and Emmons}{1990}]%
        {king1990}
\bibfield{author}{\bibinfo{person}{Laura~A. King} {and}
  \bibinfo{person}{Robert~A. Emmons}.} \bibinfo{year}{1990}\natexlab{}.
\newblock \showarticletitle{Conflict over Emotional Expression:
  {{Psychological}} and Physical Correlates.}
\newblock \bibinfo{journal}{\emph{Journal of Personality and Social
  Psychology}} \bibinfo{volume}{58}, \bibinfo{number}{5}
  (\bibinfo{year}{1990}), \bibinfo{pages}{864--877}.
\newblock
\showISSN{1939-1315, 0022-3514}


\bibitem[\protect\citeauthoryear{Kline}{Kline}{2015}]%
        {kline2015}
\bibfield{author}{\bibinfo{person}{Rex Kline}.}
  \bibinfo{year}{2015}\natexlab{}.
\newblock \bibinfo{booktitle}{\emph{Principles and Practice of Structural
  Equation Modeling} (\bibinfo{edition}{fourth} ed.)}.
\newblock \bibinfo{publisher}{{Guilford Press}}, \bibinfo{address}{New York,
  NY}.
\newblock


\bibitem[\protect\citeauthoryear{Koo and Li}{Koo and Li}{2016}]%
        {koo2016}
\bibfield{author}{\bibinfo{person}{Terry~K. Koo} {and} \bibinfo{person}{Mae~Y.
  Li}.} \bibinfo{year}{2016}\natexlab{}.
\newblock \showarticletitle{A Guideline of Selecting and Reporting Intraclass
  Correlation Coefficients for Reliability Research}.
\newblock \bibinfo{journal}{\emph{Journal of Chiropractic Medicine}}
  \bibinfo{volume}{15}, \bibinfo{number}{2} (\bibinfo{year}{2016}),
  \bibinfo{pages}{155--163}.
\newblock
\showISSN{15563707}


\bibitem[\protect\citeauthoryear{Kring, Smith, and Neale}{Kring
  et~al\mbox{.}}{1994}]%
        {kring1994}
\bibfield{author}{\bibinfo{person}{Ann~M. Kring}, \bibinfo{person}{David~A.
  Smith}, {and} \bibinfo{person}{John~M. Neale}.}
  \bibinfo{year}{1994}\natexlab{}.
\newblock \showarticletitle{Individual Differences in Dispositional
  Expressiveness: {{Development}} and Validation of the Emotional Expressivity
  Scale}.
\newblock \bibinfo{journal}{\emph{Journal of Personality and Social
  Psychology}} \bibinfo{volume}{66}, \bibinfo{number}{5}
  (\bibinfo{year}{1994}), \bibinfo{pages}{934--949}.
\newblock


\bibitem[\protect\citeauthoryear{Li and Deng}{Li and Deng}{2020}]%
        {li2020deep}
\bibfield{author}{\bibinfo{person}{Shan Li} {and} \bibinfo{person}{Weihong
  Deng}.} \bibinfo{year}{2020}\natexlab{}.
\newblock \showarticletitle{Deep Facial Expression Recognition: A Survey}.
\newblock \bibinfo{journal}{\emph{IEEE Transactions on Affective Computing}}
  (\bibinfo{year}{2020}).
\newblock


\bibitem[\protect\citeauthoryear{Lin, Girard, and Morency}{Lin
  et~al\mbox{.}}{2020}]%
        {lin2019}
\bibfield{author}{\bibinfo{person}{Victoria Lin}, \bibinfo{person}{Jeffrey~M.
  Girard}, {and} \bibinfo{person}{Louis-Philippe Morency}.}
  \bibinfo{year}{2020}\natexlab{}.
\newblock \showarticletitle{Context-Dependent Models for Predicting and
  Characterizing Facial Expressiveness}. In
  \bibinfo{booktitle}{\emph{Proceedings of the AAAI-20 Workshop on Affective
  Content Analysis}}, Vol.~\bibinfo{volume}{2614}. \bibinfo{publisher}{AAAI},
  \bibinfo{address}{New York, NY}, \bibinfo{pages}{11--28}.
\newblock


\bibitem[\protect\citeauthoryear{Littlewort, Bartlett, and Lee}{Littlewort
  et~al\mbox{.}}{2007}]%
        {littlewort2007faces}
\bibfield{author}{\bibinfo{person}{Gwen~C. Littlewort},
  \bibinfo{person}{Marian~Stewart Bartlett}, {and} \bibinfo{person}{Kang Lee}.}
  \bibinfo{year}{2007}\natexlab{}.
\newblock \showarticletitle{Faces of Pain: Automated Measurement of Spontaneous
  Facial Expressions of Genuine and Posed Pain}. In
  \bibinfo{booktitle}{\emph{Proceedings of the 9th International Conference on
  Multimodal Interfaces}}. \bibinfo{publisher}{ACM}, \bibinfo{address}{Nagoya,
  Japan}, \bibinfo{pages}{15--21}.
\newblock


\bibitem[\protect\citeauthoryear{McCrae and Costa}{McCrae and Costa}{1999}]%
        {mccrae1999}
\bibfield{author}{\bibinfo{person}{Robert~R. McCrae} {and}
  \bibinfo{person}{Paul~T. Costa}.} \bibinfo{year}{1999}\natexlab{}.
\newblock \showarticletitle{A Five-Factor Theory of Personality}.
\newblock In \bibinfo{booktitle}{\emph{Handbook of Personality: {{Theory}} and
  Research} (\bibinfo{edition}{second} ed.)},
  \bibfield{editor}{\bibinfo{person}{L.~A. Pervin} {and}
  \bibinfo{person}{O.~John}} (Eds.). \bibinfo{publisher}{{Guilford Press}},
  \bibinfo{address}{{New York, NY}}, \bibinfo{pages}{139--153}.
\newblock


\bibitem[\protect\citeauthoryear{McGraw and Wong}{McGraw and Wong}{1996}]%
        {mcgraw1996}
\bibfield{author}{\bibinfo{person}{Kenneth~O McGraw} {and} \bibinfo{person}{S~P
  Wong}.} \bibinfo{year}{1996}\natexlab{}.
\newblock \showarticletitle{Forming Inferences about Some Intraclass
  Correlation Coefficients}.
\newblock \bibinfo{journal}{\emph{Psychological Methods}} \bibinfo{volume}{1},
  \bibinfo{number}{1} (\bibinfo{year}{1996}), \bibinfo{pages}{30--46}.
\newblock


\bibitem[\protect\citeauthoryear{McNeish and Wolf}{McNeish and Wolf}{2020}]%
        {mcneish2020}
\bibfield{author}{\bibinfo{person}{Daniel McNeish} {and}
  \bibinfo{person}{Melissa~Gordon Wolf}.} \bibinfo{year}{2020}\natexlab{}.
\newblock \showarticletitle{Thinking Twice about Sum Scores}.
\newblock \bibinfo{journal}{\emph{Behavior Research Methods}}
  (\bibinfo{year}{2020}).
\newblock
\showISSN{1554-3528}


\bibitem[\protect\citeauthoryear{Merkle and Rosseel}{Merkle and
  Rosseel}{2018}]%
        {merkle2018}
\bibfield{author}{\bibinfo{person}{Edgar~C. Merkle} {and} \bibinfo{person}{Yves
  Rosseel}.} \bibinfo{year}{2018}\natexlab{}.
\newblock \showarticletitle{Blavaan: {{Bayesian}} Structural Equation Models
  via Parameter Expansion}.
\newblock \bibinfo{journal}{\emph{Journal of Statistical Software}}
  \bibinfo{volume}{85}, \bibinfo{number}{1} (\bibinfo{year}{2018}),
  \bibinfo{pages}{1--30}.
\newblock
\showISSN{1548-7660}


\bibitem[\protect\citeauthoryear{Mohammadi and Vinciarelli}{Mohammadi and
  Vinciarelli}{2012}]%
        {mohammadi2012automatic}
\bibfield{author}{\bibinfo{person}{Gelareh Mohammadi} {and}
  \bibinfo{person}{Alessandro Vinciarelli}.} \bibinfo{year}{2012}\natexlab{}.
\newblock \showarticletitle{Automatic Personality Perception: Prediction of
  Trait Attribution based on Prosodic Features}.
\newblock \bibinfo{journal}{\emph{IEEE Transactions on Affective Computing}}
  \bibinfo{volume}{3}, \bibinfo{number}{3} (\bibinfo{year}{2012}),
  \bibinfo{pages}{273--284}.
\newblock


\bibitem[\protect\citeauthoryear{Montaci{\'e} and Caraty}{Montaci{\'e} and
  Caraty}{2012}]%
        {montacie2012pitch}
\bibfield{author}{\bibinfo{person}{Claude Montaci{\'e}} {and}
  \bibinfo{person}{Marie-Jos{\'e} Caraty}.} \bibinfo{year}{2012}\natexlab{}.
\newblock \showarticletitle{Pitch and Intonation Contribution to Speakers'
  Traits Classification}. In \bibinfo{booktitle}{\emph{Proceedings of the 13th
  Annual Conference of the International Speech Communication Association}}.
  \bibinfo{publisher}{ISCA}, \bibinfo{address}{Portland, OR},
  \bibinfo{pages}{526--529}.
\newblock


\bibitem[\protect\citeauthoryear{Pasupathi, Carstensen, Levenson, and
  Gottman}{Pasupathi et~al\mbox{.}}{1999}]%
        {pasupathi1999responsive}
\bibfield{author}{\bibinfo{person}{Monisha Pasupathi},
  \bibinfo{person}{Laura~L. Carstensen}, \bibinfo{person}{Robert~W. Levenson},
  {and} \bibinfo{person}{John~M. Gottman}.} \bibinfo{year}{1999}\natexlab{}.
\newblock \showarticletitle{Responsive listening in long-married couples: A
  psycholinguistic perspective}.
\newblock \bibinfo{journal}{\emph{Journal of Nonverbal Behavior}}
  \bibinfo{volume}{23}, \bibinfo{number}{2} (\bibinfo{year}{1999}),
  \bibinfo{pages}{173--193}.
\newblock


\bibitem[\protect\citeauthoryear{Pennebaker, Boyd, Jordan, and
  Blackburn}{Pennebaker et~al\mbox{.}}{2015}]%
        {pennebaker2015}
\bibfield{author}{\bibinfo{person}{James~W. Pennebaker},
  \bibinfo{person}{Ryan~L. Boyd}, \bibinfo{person}{Kayla Jordan}, {and}
  \bibinfo{person}{Kate Blackburn}.} \bibinfo{year}{2015}\natexlab{}.
\newblock \bibinfo{booktitle}{\emph{The Development and Psychometric Properties
  of {{LIWC2015}}}}.
\newblock \bibinfo{publisher}{{University of Texas at Austin}},
  \bibinfo{address}{{Austin, TX, USA}}.
\newblock


\bibitem[\protect\citeauthoryear{Petridis and Pantic}{Petridis and
  Pantic}{2008}]%
        {petridis2008audiovisual}
\bibfield{author}{\bibinfo{person}{Stavros Petridis} {and}
  \bibinfo{person}{Maja Pantic}.} \bibinfo{year}{2008}\natexlab{}.
\newblock \showarticletitle{Audiovisual Discrimination between Laughter and
  Speech}. In \bibinfo{booktitle}{\emph{Proceedings of the 33rd IEEE
  International Conference on Acoustics, Speech and Signal Processing}}.
  \bibinfo{publisher}{IEEE}, \bibinfo{address}{Las Vegas, NV},
  \bibinfo{pages}{5117--5120}.
\newblock


\bibitem[\protect\citeauthoryear{Polzehl, Moller, and Metze}{Polzehl
  et~al\mbox{.}}{2010}]%
        {polzehl2010automatically}
\bibfield{author}{\bibinfo{person}{Tim Polzehl}, \bibinfo{person}{Sebastian
  Moller}, {and} \bibinfo{person}{Florian Metze}.}
  \bibinfo{year}{2010}\natexlab{}.
\newblock \showarticletitle{Automatically Assessing Personality from Speech}.
  In \bibinfo{booktitle}{\emph{Proceedings of the 4th IEEE International
  Conference on Semantic Computing}}. \bibinfo{publisher}{IEEE},
  \bibinfo{address}{Pittsburgh, PA}, \bibinfo{pages}{134--140}.
\newblock


\bibitem[\protect\citeauthoryear{Ponce-L{\'o}pez, Chen, Oliu, Corneanu,
  Clap{\'e}s, Guyon, Bar{\'o}, Escalante, and Escalera}{Ponce-L{\'o}pez
  et~al\mbox{.}}{2016}]%
        {ponce2016chalearn}
\bibfield{author}{\bibinfo{person}{V{\'\i}ctor Ponce-L{\'o}pez},
  \bibinfo{person}{Baiyu Chen}, \bibinfo{person}{Marc Oliu},
  \bibinfo{person}{Ciprian Corneanu}, \bibinfo{person}{Albert Clap{\'e}s},
  \bibinfo{person}{Isabelle Guyon}, \bibinfo{person}{Xavier Bar{\'o}},
  \bibinfo{person}{Hugo~Jair Escalante}, {and} \bibinfo{person}{Sergio
  Escalera}.} \bibinfo{year}{2016}\natexlab{}.
\newblock \showarticletitle{Chalearn lap 2016: First round challenge on first
  impressions-dataset and results}. In \bibinfo{booktitle}{\emph{European
  Conference on Computer Vision}}. \bibinfo{publisher}{Springer},
  \bibinfo{address}{Amsterdam, the Netherlands}, \bibinfo{pages}{400--418}.
\newblock


\bibitem[\protect\citeauthoryear{Rosenthal}{Rosenthal}{1982}]%
        {rosenthal1982}
\bibfield{author}{\bibinfo{person}{Robert Rosenthal}.}
  \bibinfo{year}{1982}\natexlab{}.
\newblock \showarticletitle{Conducting Judgment Studies}.
\newblock In \bibinfo{booktitle}{\emph{Handbook of Methods in Nonverbal
  Behavior Research}}, \bibfield{editor}{\bibinfo{person}{Klaus~R. Scherer}
  {and} \bibinfo{person}{Paul Ekman}} (Eds.). \bibinfo{publisher}{{Cambridge
  University Press}}, \bibinfo{address}{{New York, NY}},
  \bibinfo{pages}{287--365}.
\newblock


\bibitem[\protect\citeauthoryear{Sayette, Creswell, Dimoff, Fairbairn, Cohn,
  Heckman, Kirchner, Levine, and Moreland}{Sayette et~al\mbox{.}}{2012}]%
        {sayette2012}
\bibfield{author}{\bibinfo{person}{Michael~A. Sayette},
  \bibinfo{person}{Kasey~G. Creswell}, \bibinfo{person}{John~D. Dimoff},
  \bibinfo{person}{Catharine~E. Fairbairn}, \bibinfo{person}{Jeffrey~F. Cohn},
  \bibinfo{person}{Bryan~W. Heckman}, \bibinfo{person}{Thomas~R. Kirchner},
  \bibinfo{person}{John~M. Levine}, {and} \bibinfo{person}{Richard~L.
  Moreland}.} \bibinfo{year}{2012}\natexlab{}.
\newblock \showarticletitle{Alcohol and Group Formation: {{A}} Multimodal
  Investigation of the Effects of Alcohol on Emotion and Social Bonding}.
\newblock \bibinfo{journal}{\emph{Psychological Science}} \bibinfo{volume}{23},
  \bibinfo{number}{8} (\bibinfo{year}{2012}), \bibinfo{pages}{869--878}.
\newblock
\showISSN{0956-7976}


\bibitem[\protect\citeauthoryear{Schuller, M{\"u}ller, H{\"o}rnler,
  H{\"o}thker, Konosu, and Rigoll}{Schuller et~al\mbox{.}}{2007}]%
        {schuller2007audiovisual}
\bibfield{author}{\bibinfo{person}{Bj{\"o}rn Schuller}, \bibinfo{person}{Ronald
  M{\"u}ller}, \bibinfo{person}{Benedikt H{\"o}rnler}, \bibinfo{person}{Anja
  H{\"o}thker}, \bibinfo{person}{Hitoshi Konosu}, {and}
  \bibinfo{person}{Gerhard Rigoll}.} \bibinfo{year}{2007}\natexlab{}.
\newblock \showarticletitle{Audiovisual Recognition of Spontaneous Interest
  within Conversations}. In \bibinfo{booktitle}{\emph{Proceedings of the 9th
  International Conference on Multimodal interfaces}}.
  \bibinfo{publisher}{ACM}, \bibinfo{address}{Nagoya, Japan},
  \bibinfo{pages}{30--37}.
\newblock


\bibitem[\protect\citeauthoryear{Schuller, Steidl, Batliner, N{\"o}th,
  Vinciarelli, Burkhardt, Son, Weninger, Eyben, Bocklet,
  et~al\mbox{.}}{Schuller et~al\mbox{.}}{2012}]%
        {schuller2012interspeech}
\bibfield{author}{\bibinfo{person}{Bj{\"o}rn Schuller}, \bibinfo{person}{Stefan
  Steidl}, \bibinfo{person}{Anton Batliner}, \bibinfo{person}{Elmar N{\"o}th},
  \bibinfo{person}{Alessandro Vinciarelli}, \bibinfo{person}{Felix Burkhardt},
  \bibinfo{person}{Rob~van Son}, \bibinfo{person}{Felix Weninger},
  \bibinfo{person}{Florian Eyben}, \bibinfo{person}{Tobias Bocklet},
  {et~al\mbox{.}}} \bibinfo{year}{2012}\natexlab{}.
\newblock \showarticletitle{The InterSpeech 2012 Speaker Trait Challenge}. In
  \bibinfo{booktitle}{\emph{Thirteenth Annual Conference of the International
  Speech Communication Association}}. \bibinfo{publisher}{ISCA},
  \bibinfo{address}{Portland, OR}, \bibinfo{pages}{254--257}.
\newblock


\bibitem[\protect\citeauthoryear{Sharif~Razavian, Azizpour, Sullivan, and
  Carlsson}{Sharif~Razavian et~al\mbox{.}}{2014}]%
        {sharif2014cnn}
\bibfield{author}{\bibinfo{person}{Ali Sharif~Razavian},
  \bibinfo{person}{Hossein Azizpour}, \bibinfo{person}{Josephine Sullivan},
  {and} \bibinfo{person}{Stefan Carlsson}.} \bibinfo{year}{2014}\natexlab{}.
\newblock \showarticletitle{CNN Features Off-the-Shelf: An Astounding Baseline
  for Recognition}. In \bibinfo{booktitle}{\emph{Proceedings of the IEEE
  Conference on Computer Vision and Pattern Recognition Workshops}}.
  \bibinfo{publisher}{IEEE}, \bibinfo{address}{Columbus, OH},
  \bibinfo{pages}{806--813}.
\newblock


\bibitem[\protect\citeauthoryear{Staiano, Lepri, Subramanian, Sebe, and
  Pianesi}{Staiano et~al\mbox{.}}{2011}]%
        {staiano2011automatic}
\bibfield{author}{\bibinfo{person}{Jacopo Staiano}, \bibinfo{person}{Bruno
  Lepri}, \bibinfo{person}{Ramanathan Subramanian}, \bibinfo{person}{Nicu
  Sebe}, {and} \bibinfo{person}{Fabio Pianesi}.}
  \bibinfo{year}{2011}\natexlab{}.
\newblock \showarticletitle{Automatic modeling of personality states in small
  group interactions}. In \bibinfo{booktitle}{\emph{Proceedings of the 19th ACM
  International Conference on Multimedia}}. \bibinfo{publisher}{ACM},
  \bibinfo{address}{Scottsdale, AZ}, \bibinfo{pages}{989--992}.
\newblock


\bibitem[\protect\citeauthoryear{Sullivan, Leifker, and Marshall}{Sullivan
  et~al\mbox{.}}{2018}]%
        {sullivan2018}
\bibfield{author}{\bibinfo{person}{Timothy~J. Sullivan},
  \bibinfo{person}{Feea~R. Leifker}, {and} \bibinfo{person}{Amy~D. Marshall}.}
  \bibinfo{year}{2018}\natexlab{}.
\newblock \showarticletitle{Observed Emotional Expressivity, Posttraumatic
  Stress Disorder Symptoms, and Intimate Partner Violence Perpetration among
  Community Couples}.
\newblock \bibinfo{journal}{\emph{Journal of Traumatic Stress}}
  \bibinfo{volume}{31}, \bibinfo{number}{3} (\bibinfo{year}{2018}),
  \bibinfo{pages}{352--361}.
\newblock
\showISSN{1573-6598}


\bibitem[\protect\citeauthoryear{Tellegen, Lykken, Bouchard, Wilcox, Segal, and
  Rich}{Tellegen et~al\mbox{.}}{1988}]%
        {tellegen1988}
\bibfield{author}{\bibinfo{person}{Auke Tellegen}, \bibinfo{person}{David~T.
  Lykken}, \bibinfo{person}{Thomas~J. Bouchard}, \bibinfo{person}{Kimerly~J.
  Wilcox}, \bibinfo{person}{Nancy~L. Segal}, {and} \bibinfo{person}{Stephen
  Rich}.} \bibinfo{year}{1988}\natexlab{}.
\newblock \showarticletitle{Personality Similarity in Twins Reared Apart and
  Together}.
\newblock \bibinfo{journal}{\emph{Journal of Personality and Social
  Psychology}} \bibinfo{volume}{54}, \bibinfo{number}{6}
  (\bibinfo{year}{1988}), \bibinfo{pages}{1031--1039}.
\newblock


\bibitem[\protect\citeauthoryear{Tong, Liao, and Ji}{Tong
  et~al\mbox{.}}{2007}]%
        {tong2007facial}
\bibfield{author}{\bibinfo{person}{Yan Tong}, \bibinfo{person}{Wenhui Liao},
  {and} \bibinfo{person}{Qiang Ji}.} \bibinfo{year}{2007}\natexlab{}.
\newblock \showarticletitle{Facial Action Unit Recognition by Exploiting their
  Dynamic and Semantic Relationships}.
\newblock \bibinfo{journal}{\emph{IEEE Transactions on Pattern Analysis and
  Machine Intelligence}} \bibinfo{volume}{29}, \bibinfo{number}{10}
  (\bibinfo{year}{2007}), \bibinfo{pages}{1683--1699}.
\newblock


\bibitem[\protect\citeauthoryear{Tzirakis, Trigeorgis, Nicolaou, Schuller, and
  Zafeiriou}{Tzirakis et~al\mbox{.}}{2017}]%
        {tzirakis2017end}
\bibfield{author}{\bibinfo{person}{Panagiotis Tzirakis},
  \bibinfo{person}{George Trigeorgis}, \bibinfo{person}{Mihalis~A. Nicolaou},
  \bibinfo{person}{Bj{\"o}rn~W. Schuller}, {and} \bibinfo{person}{Stefanos
  Zafeiriou}.} \bibinfo{year}{2017}\natexlab{}.
\newblock \showarticletitle{End-to-End Multimodal Emotion Recognition using
  Deep Neural Networks}.
\newblock \bibinfo{journal}{\emph{IEEE Journal of Selected Topics in Signal
  Processing}} \bibinfo{volume}{11}, \bibinfo{number}{8}
  (\bibinfo{year}{2017}), \bibinfo{pages}{1301--1309}.
\newblock


\bibitem[\protect\citeauthoryear{Valente, Kim, and Motlicek}{Valente
  et~al\mbox{.}}{2012}]%
        {valente2012annotation}
\bibfield{author}{\bibinfo{person}{Fabio Valente}, \bibinfo{person}{Samuel
  Kim}, {and} \bibinfo{person}{Petr Motlicek}.}
  \bibinfo{year}{2012}\natexlab{}.
\newblock \showarticletitle{Annotation and Recognition of Personality Traits in
  Spoken Conversations from the AMI Meetings Corpus}. In
  \bibinfo{booktitle}{\emph{Proceedings of the 13th Annual Conference of the
  International Speech Communication Association}}. \bibinfo{publisher}{ISCA},
  \bibinfo{address}{Portland, OR}, \bibinfo{pages}{1183--1186}.
\newblock


\bibitem[\protect\citeauthoryear{Valstar, Schuller, Krajewski, Cowie, and
  Pantic}{Valstar et~al\mbox{.}}{2014}]%
        {valstar2014}
\bibfield{author}{\bibinfo{person}{Michel~F. Valstar},
  \bibinfo{person}{Bj{\"o}rn~W. Schuller}, \bibinfo{person}{Jarek Krajewski},
  \bibinfo{person}{Roddy Cowie}, {and} \bibinfo{person}{Maja Pantic}.}
  \bibinfo{year}{2014}\natexlab{}.
\newblock \showarticletitle{{{AVEC}} 2014: The 4th International Audio/Visual
  Emotion Challenge and Workshop}. In \bibinfo{booktitle}{\emph{Proceedings of
  the ACM International Conference on Multimedia}}. \bibinfo{publisher}{ACM},
  \bibinfo{address}{Orlando, FL}, \bibinfo{pages}{1243--1244}.
\newblock


\bibitem[\protect\citeauthoryear{Vidrascu and Devillers}{Vidrascu and
  Devillers}{2005}]%
        {vidrascu2005detection}
\bibfield{author}{\bibinfo{person}{Laurence Vidrascu} {and}
  \bibinfo{person}{Laurence Devillers}.} \bibinfo{year}{2005}\natexlab{}.
\newblock \showarticletitle{Detection of Real-Life Emotions in Call Centers}.
  In \bibinfo{booktitle}{\emph{Proceedings of the 9th European Conference on
  Speech Communication and Technology}}. \bibinfo{publisher}{ISCA},
  \bibinfo{address}{Lisbon, Portugal}, \bibinfo{pages}{1841--1844}.
\newblock


\bibitem[\protect\citeauthoryear{Vinciarelli and Mohammadi}{Vinciarelli and
  Mohammadi}{2014}]%
        {vinciarelli2014}
\bibfield{author}{\bibinfo{person}{Alessandro Vinciarelli} {and}
  \bibinfo{person}{Gelareh Mohammadi}.} \bibinfo{year}{2014}\natexlab{}.
\newblock \showarticletitle{A Survey of Personality Computing}.
\newblock \bibinfo{journal}{\emph{IEEE Transactions on Affective Computing}}
  \bibinfo{volume}{5}, \bibinfo{number}{3} (\bibinfo{year}{2014}),
  \bibinfo{pages}{273--291}.
\newblock
\showISSN{1949-3045}


\bibitem[\protect\citeauthoryear{Wright}{Wright}{2014}]%
        {wright2014b}
\bibfield{author}{\bibinfo{person}{Aidan G.~C. Wright}.}
  \bibinfo{year}{2014}\natexlab{}.
\newblock \showarticletitle{Current Directions in Personality Science and the
  Potential for Advances through Computing}.
\newblock \bibinfo{journal}{\emph{IEEE Transactions on Affective Computing}}
  \bibinfo{volume}{5}, \bibinfo{number}{3} (\bibinfo{year}{2014}),
  \bibinfo{pages}{292--296}.
\newblock


\bibitem[\protect\citeauthoryear{Zeng, Hu, Roisman, Wen, Fu, and Huang}{Zeng
  et~al\mbox{.}}{2007}]%
        {zeng2007audio}
\bibfield{author}{\bibinfo{person}{Zhihong Zeng}, \bibinfo{person}{Yuxiao Hu},
  \bibinfo{person}{Glenn~I. Roisman}, \bibinfo{person}{Zhen Wen},
  \bibinfo{person}{Yun Fu}, {and} \bibinfo{person}{Thomas~S. Huang}.}
  \bibinfo{year}{2007}\natexlab{}.
\newblock \showarticletitle{Audio-visual Spontaneous Emotion Recognition}.
\newblock In \bibinfo{booktitle}{\emph{Artifical Intelligence for Human
  Computing}}. \bibinfo{publisher}{Springer}, \bibinfo{address}{New York, NY},
  \bibinfo{pages}{72--90}.
\newblock


\bibitem[\protect\citeauthoryear{Zeng, Pantic, Roisman, and Huang}{Zeng
  et~al\mbox{.}}{2009}]%
        {zeng2009}
\bibfield{author}{\bibinfo{person}{Zhihong Zeng}, \bibinfo{person}{Maja
  Pantic}, \bibinfo{person}{Glenn~I. Roisman}, {and} \bibinfo{person}{Thomas~S.
  Huang}.} \bibinfo{year}{2009}\natexlab{}.
\newblock \showarticletitle{A Survey of Affect Recognition Methods: Audio,
  Visual, and Spontaneous Expressions}.
\newblock \bibinfo{journal}{\emph{IEEE Transactions on Pattern Analysis and
  Machine Intelligence}} \bibinfo{volume}{31}, \bibinfo{number}{1}
  (\bibinfo{year}{2009}), \bibinfo{pages}{39--58}.
\newblock


\bibitem[\protect\citeauthoryear{Zou and Hastie}{Zou and Hastie}{2005}]%
        {zou2005}
\bibfield{author}{\bibinfo{person}{Hui Zou} {and} \bibinfo{person}{Trevor
  Hastie}.} \bibinfo{year}{2005}\natexlab{}.
\newblock \showarticletitle{Regularization and Variable Selection via the
  Elastic Net}.
\newblock \bibinfo{journal}{\emph{Journal of the Royal Statistical Society:
  Series B (Statistical Methodology)}} \bibinfo{volume}{67},
  \bibinfo{number}{2} (\bibinfo{year}{2005}), \bibinfo{pages}{301--320}.
\newblock
\showISSN{1369-7412, 1467-9868}


\end{thebibliography}
\balance 

\end{document}